\documentclass[10pt, journal,a4paper,final,oneside,twocolumn]{IEEEtran}
\usepackage{mathrsfs}
\usepackage{pifont}
\usepackage{bbding}
\usepackage{tipa}
\usepackage{cite}
\usepackage{epsfig}
\usepackage{graphicx}
\usepackage{url}
\usepackage{amsfonts}
\usepackage{multicol}
\usepackage{float}
\usepackage{times}
\usepackage{cite}
\usepackage{psfrag}
\usepackage{subfigure}
\usepackage{stfloats}
\usepackage{amsmath}
\usepackage{footnote}
\usepackage{array}
\usepackage{amsmath,epsfig}
\usepackage{stmaryrd}
\usepackage{amssymb}
\usepackage{epic}
\usepackage{graphicx}
\usepackage{curves}
\usepackage{color}
\usepackage{balance}

\begin{document}
\newtheorem{theorem}{Theorem}
\newtheorem{proposition}{Proposition}
\newtheorem{definition}{Definition}
\newtheorem{lemma}{Lemma}
\newtheorem{corollary}{Corollary}
\newtheorem{remark}{Remark}
\newtheorem{construction}{Construction}
\newtheorem{algorithm}{Algorithm}

\newcommand{\supp}{\mathop{\rm supp}}
\newcommand{\sinc}{\mathop{\rm sinc}}
\newcommand{\spann}{\mathop{\rm span}}
\newcommand{\essinf}{\mathop{\rm ess\,inf}}
\newcommand{\esssup}{\mathop{\rm ess\,sup}}
\newcommand{\Lip}{\rm Lip}
\newcommand{\sign}{\mathop{\rm sign}}
\newcommand{\osc}{\mathop{\rm osc}}
\newcommand{\R}{{\mathbb{R}}}
\newcommand{\Z}{{\mathbb{Z}}}
\newcommand{\C}{{\mathbb{C}}}
\title{Distributed Caching for Data Dissemination in the Downlink of Heterogeneous Networks}
\author{{Jun~Li~\emph{Member, IEEE},~Youjia Chen,~Zihuai Lin~\emph{Senior Member, IEEE},\\~Wen Chen~\emph{Senior Member, IEEE},~Branka Vucetic~\emph{Fellow, IEEE}, and Lajos Hanzo~\emph{Fellow, IEEE}}
\thanks{Jun~Li is with the School of Electronic and Optical Engineering, Nanjing University of Science and Technology, Nanjing, 210094, CHINA. Email: jleesr80@gmail.com.}
\thanks{Youjia Chen, Zihuai Lin, and Branka Vucetic are with School of Electrical and Information Engineering, The University of Sydney, NSW, 2006, AUSTRALIA. E-mail: \{youjia.chen,zihuai.lin,branka.vucetic\}@sydney.edu.au.}
\thanks{Wen Chen is with the Department of Electronic Engineering, Shanghai Jiao Tong University, CHINA. E-mail: wenchen@sjtu.edu.cn.}
\thanks{Lajos Hanzo is with  Department of Electronics and Computer Science, University of Southampton, U.K. E-mail: lh@ecs.soton.ac.uk.}
\thanks{This work was supported by Australian Research Council Programs $DP120100405$, by National 973 Project $2012CB316106$, by NSF China $61328101$, $61271230$ and $61472190$, by STCSM Science and Technology Innovation Program $13510711200$, and by SEU National Key Lab on Mobile Communications $2013D11$ and $2013D02$.}}

\markboth{IEEE Transactions on Communications, Accepted for Publication}
{Li \MakeLowercase{\textit{et al.}}: Distributed Caching for Data Dissemination in the Downlink of Heterogeneous Networks}

\maketitle
\begin{abstract}
Heterogeneous cellular networks (HCN) with embedded small cells are considered, where multiple mobile users wish to download network content of different popularity. By caching data into the small-cell base stations (SBS), we will design distributed caching optimization algorithms via belief propagation (BP) for minimizing the downloading latency. First, we derive the delay-minimization objective function (OF) and formulate an optimization problem. Then we develop a framework for modeling the underlying HCN topology with the aid of a factor graph. Furthermore, distributed BP algorithm is proposed based on the network's factor graph. Next, we prove that a fixed point of convergence exists for our distributed BP algorithm. In order to reduce the complexity of the BP, we propose a heuristic BP algorithm. Furthermore, we evaluate the average downloading performance of our HCN for different numbers and locations of the base stations (BS) and mobile users (MU), with the aid of stochastic geometry theory. By modeling the nodes distributions using a Poisson point process, we develop the expressions of the average factor graph degree distribution, as well as an upper bound of the outage probability for random caching schemes. We also improve the performance of random caching. Our simulations show that (1) the proposed distributed BP algorithm has a near-optimal delay performance, approaching that of the high-complexity exhaustive search method, (2) the modified BP offers a good delay performance at a low communication complexity, (3) both the average degree distribution and the outage upper bound analysis relying on stochastic geometry match well with our Monte-Carlo simulations, and (4) the optimization based on the upper bound provides both a better outage and a better delay performance than the benchmarks.
\end{abstract}

\section{Introduction}\label{sec:introduction}
Wireless data traffic is expected to increase by a factor of $40$ over the next five years, from the current level of $93$ Petabytes to $3600$ Petabytes per month~\cite{Cisco:Online}, driven by a rapid increase in the number of mobile users (MU) and aggravated by their bandwidth-hungry mobile applications. A promising approach to enhancing the network capacity is to embed small cells relying on low-power base stations (BS) into the existing macro-cell based networks. These networks, which are referred to as heterogeneous cellular networks (HCN)~\cite{Boccardi:ICM14,Damnjanovic:IWC11,Hanzo:10,Bayat:TCOM14,Mirahmadi:TCOM14,Gupta:TCOM14}, typically contain regularly deployed macro-cells and embedded femto-cells as well as pico-cells~\cite{Kishiyama:IWC13,Nakamura:ICM13,YunLi:WCI13} that are served by macro-cell BSs (MBS) and small-cell BSs (SBS), respectively. The aim of these flexibly deployed low-power SBSs is to eliminate the coverage holes and to increase the capacity in hot-spots.

There is evidence that the MUs' downloading of video on-demand files is the main reason for the growth of data traffic over cellular networks~\cite{Golrezaei:CM13}. According to the prediction of Cisco on mobile data traffic, the mobile video streaming traffic will occupy $72\%$ percentage of the overall mobile data traffic by $2019$. Often, there are numerous repetitive downloading requests of popular contents, such as online blockbusters, leading to redundant data streaming. The redundancy of data transmissions can be reduced by locally storing popular data, known as caching, into the local SBSs, effectively forming a local cloud caching system (LCCS). The LCCS brings the content closer to the MUs and alleviates redundant data transmissions via redirecting the downloading requests to local SBSs. Also, the SBSs are willing to cache files into their buffers as long as they can, since caching is capable of significantly reducing the tele-traffic load on their back-haul channels, which are expensive.

In~\cite{YongLi:TITS14}, the authors study the caching strategies of delay-tolerant vehicular networks, where the data subscribers and ``helpers'' are always moving and the links between them are opportunistic. By proposing an efficient algorithm to carefully allocate the network resources to mobile data, the decision is made as to which content should use the erasure coding, as well as conceiving the coding policy for each mobile data. In~\cite{Jianliang:TKDE04}, optimal cache replacement policies are investigated. The cache replacement process takes place after the data caching process has been completed, and determines which particular data item should be deleted from the cache, when the available storage space is insufficient for accommodating an item to be cached.

Since the HCN structure has been widely adopted in current cellular networks and will prevail in near-future networks, we are interested in the SBS-based LCCS in the context of HCNs. In contrast to the vehicular networks discussed in~\cite{YongLi:TITS14,Li:TMC13}, where the mobility and the opportunistic communication contact are important issues, in the context of HCNs, the BSs are always fixed, and the MUs are assumed to be moving at a low speed. Thus, we ignore the mobility issues in the HCNs and assume that each MU is associated with a fixed BS during file-downloading. At the time of writing, there are already technical reports highlighting the advantages of caching in HCNs~\cite{David:TR13,Haig:TR14,Intel:TR12}. Based on these reports, the LCCS with SBS caching for HCNs is capable of efficiently 1) reducing the transmission latency due to short distance between the SBSs and the MUs, 2) offloading redundant data streams from MBSs, and 3) alleviating heavy burdens on the back-haul channels of the SBSs. Therefore, SBS-based caching will bring about significant breakthroughs for future HCNs.

The concept of caching is common in wireline networks and computer systems. However, research on efficient caching design for wireless cellular networks relying on small cells is still in its infancy~\cite{Golrezaei:CM13,Xiaofei:CM14}. Usually, data caching consists of two phases: data placement and data transmission. During the data placement phase, data is cached into local SBSs in order to form an LCCS. In the data transmission phase, MUs request data from the LCCS. The focus of wireless caching research is mainly on the optimization of data placement for ensuring that the downloading latency is minimized. The caching optimization is a non-trivial problem. This is due to the massive scale of video contents to be stored in the limited memory of the SBSs.

The survey papers~\cite{Golrezaei:CM13,Xiaofei:CM14} report on a range of attractive caching architectures conceived for future cellular networks. In~\cite{Golrezaei:ICC12}, a caching scheme is proposed for a device-to-device (D2D) based cellular network on the MUs' caching of popular data. In this scheme, the D2D cluster size was optimized for reducing the downloading delay. In~\cite{Ji:arXiv1305,Mingyue:ISIT13}, the authors propose a caching scheme for wireless sensor networks, where the protocol model of~\cite{Gupta:IT00} is adopted. In~\cite{Shuan:TIT13}, a femto-caching scheme is proposed for a cellular network combined with SBSs, where the data placement at the SBSs is optimized in a centralized manner for reducing the transmission delay imposed. However,~\cite{Shuan:TIT13} considers an idealized system, where neither the interference nor the impact of wireless channels is taken into account. The associations between the MUs and the SBSs are pre-determined without considering the specific channel conditions encountered. Furthermore, this centralized optimization method assumes that the MBS has perfect knowledge of all the channel state information (CSI) between the MUs and SBSs, which is impractical.

Against this background, in this paper, we consider distributed caching solutions for HCNs operating under more practical considerations. Our contributions consist of two parts.
\begin{enumerate}
\begin{item}
In the first part, we propose distributed caching algorithms for enhancing the downloading performance via belief propagation (BP)~\cite{Moallem:INFO11}. The BP algorithm is capable of decomposing a global optimization problem into multiple sub-problems, thereby offering an efficient distributive approach of solving the global optimization problem~\cite{Kschischang:FGSP01,Bavarian:JSAC08,Illsoo:TWC11}. As the BP method has been widely adopted for distributively solving resource allocation in cellular networks, we arrange file placement via BP algorithms by viewing files as a type of resource.
\end{item}
\begin{item}
In the second part, we analyze the average caching performance based on stochastic geometry theory~\cite{Stoyan:book,Haenggi:JSAC09}. We are interested in optimizing the average performance of a set of HCNs, where the channels exhibit Rayleigh fading and the distributions of network nodes obey a Poisson point process (PPP)~\cite{Daley:book}.
\end{item}
\end{enumerate}

Specifically, our contributions in the first part are follows.
\begin{enumerate}
\item{We commence by deriving the delay as our optimization objective function (OF) and formulate the problem as optimizing the file placement.}
\item{We develop a framework for modeling the associated factor graph based on the topology of the network. A distributed BP algorithm is proposed based on the factor graph, which allows the file placement to be optimized in a distributed manner between the MUs and SBSs.}
\item{We prove that a fixed point exists in the proposed BP algorithm and show that the BP algorithm is capable of converging to this fixed point under certain conditions.}
\item{To reduce the communication complexity, we propose a heuristic BP algorithm.}
\end{enumerate}

Our contributions in the second part are follows.
\begin{enumerate}
\item{By following the stochastic geometry framework, we model the MUs and SBSs in the HCN as different ties of a PPP. Furthermore, we develop the average degree distribution of the factor graph in the BP algorithm.}
\item{A random caching scheme is proposed, where each SBS will cache a file with a pre-determined probability. We can characterize the average downloading performance by outage probability (OP) and develop a tight upper bound of the OP expression with a closed form under the random caching scheme.}
\item{Based on the upper bound derived, we further improve the OP performance of random caching by optimizing the probabilities for caching different files.}
\end{enumerate}

In the simulations, we first investigate the average degree distribution of the factor graph, as well as the OP and the delay of the random caching schemes, in conjunction with various PPP parameters and power settings. It is shown that both the degree distribution and our upper bound analysis match well with the results of Monte-Carlo simulations. Furthermore, the optimization based on the upper bound provides both a better OP and a better delay than the benchmarks. Then we evaluate the distributed BP algorithm in our HCNs having a fixed number of BSs and MUs. It is shown that the proposed distributed BP algorithm has a near-optimal performance, approaching that of the exhaustive search method. The heuristic BP also offers a relatively good performance, despite its significantly reduced communication complexity.

The rest of this paper is organized as follows. We describe the system model in Section~\ref{sec:system_model} and present the distributed file downloading problem relying on caching in Section~\ref{sec:caching_protocol}. We then propose a distributed BP algorithm in Section~\ref{sec:cache_optimizations_BP}, where the proof of existence for a fixed point is also presented. In Section~\ref{sec:heuristic_BP}, a heuristic BP algorithm is proposed for reducing the associated communication complexity. Our stochastic geometry based analysis is detailed in Section~\ref{sec:stochastic}, where the average degree distribution of the factor graph and the OP of the random caching scheme are developed. Our simulation results are summarized in Section~\ref{sec:simulations}, while our conclusions are provided in Section~\ref{sec:conclusions}.
\section{System Model}\label{sec:system_model}
Let us consider an HCN consisting of a single MBS and $K$ SBSs illuminating both femto-cells and pico-cells, while supporting $J$ MUs randomly located in the network. Let us denote by $\mathcal B_0$ the MBS and by $\boldsymbol {\mathcal B}=\{\mathcal B_1,\mathcal B_2,\cdots,\mathcal B_K\}$ the set of the SBSs, where $\mathcal B_k$, $k\in\boldsymbol{\mathcal K}=\{1,2,\cdots,K\}$, represents the $k$-th SBS. Furthermore, denote by $\boldsymbol{\mathcal U}=\{\mathcal U_1,\mathcal U_2,\cdots\,\mathcal U_J\}$ the set of the MUs, where $\mathcal U_j$, $j\in\boldsymbol{\mathcal J}=\{1,2,\cdots,J\}$, represents the $j$-th MU. The MBS $\mathcal B_0$ caches files into the memories of the SBSs during off-peak time via back-haul channels. Once the caching process is completed, the MBSs and SBSs are ready to act upon the downloading requests of the MUs.

We assume that a dedicated frequency band of bandwidth $W$ is allocated to the downlink channels spanning from the SBSs to the MUs for file-dissemination. For reasons of careful load balancing, we consider the ``SBS-first'' constraint, where each MU will try to download data from its adjacent SBSs, unless the required files cannot be found in these SBSs. In this case, the MU will turn to the MBS for retrieving the required files. For the sake of simplicity, we assume that the MBS will support a fixed download rate, denoted by $C_0$, for the MUs in the channels which are orthogonal to those spanning from the SBSs to MUs.

In order to satisfy the ``SBS-first'' constraint for offloading data from the MBS, some incentives may be provided for the MUs. For example, downloading from the SBSs is much cheaper than from the MBS. Here, we assume that the download rate $C_0$ supported by the MBS is never higher than the lowest download rate supported by the SBSs. This limit imposed on the download rate from the MBS will not only encourage the MUs to download from the SBSs first, but also effectively control the data traffic of the MBS imposed by file downloading.

Denote by $P_k$ the transmission power of the $k$-th SBS, and by $\sigma^2$ the noise power at each MU. The path-loss between $\mathcal B_k$ and the MU $\mathcal U_j$ is modeled as $d_{k,j}^{-\alpha}$, where $d_{k,j}$ is the distance between $\mathcal B_k$ and $\mathcal U_j$, and $\alpha$ is the path-loss exponent. The random channel between $\mathcal B_k$ and $\mathcal U_j$ is Rayleigh fading, whose coefficient $h_{k,j}$ has the average power of one. We assume that all the downlink channels spanning from the SBSs to the MUs are independent and identically distributed (\emph{i.i.d.}).

Suppose that each file is split into multiple chunks and each chunk can be downloaded by an MU in a short time slot. Due to the short downloading time of a chunk, we assume furthermore that the probability of having two MUs streaming a chunk at the same time (or within a relative delay of a few seconds) from the same SBS is basically zero~\cite{Ji:arXiv1305}. Hence, neither direct multicasting by exploiting the broadcast nature of the wireless medium nor network coding is considered. Furthermore, we focus our attention on the saturated scenario, where the SBSs keep transmitting data to the MUs~\cite{Dhillon:JSAC12}. Hence, each MU is subject to the interference imposed by all the other SBSs in $\boldsymbol {\mathcal B}$, when downloading files from its associated SBS. Given a channel realization $\textbf h_j=[h_{1,j},\cdots,h_{K,j}]$, the channel capacity between $\mathcal B_k$ and $\mathcal U_j$ can be calculated based on the signal-to-interference-plus-noise ratio (SINR) as
\begin{equation}\label{equ:SINR_capacity}
C_{k,j}=W\log\left(1+\frac{h_{k,j}^2d_{k,j}^{-\alpha}P_{k}}{\underset{{q\in{\boldsymbol{\mathcal K}}\backslash\{k\}}}{\sum}h_{q,j}^2d_{q,j}^{-\alpha}P_q+\sigma^2}\right).
\end{equation}
Due to the `SBS-first' constraint, we have $C_0\le C_{k,j}$, $\forall k\in{\boldsymbol {\mathcal K}},j\in{\boldsymbol {\mathcal J}}$.

Denote by $\boldsymbol{\mathcal F}$ the library or set of files, which consists of $Q$ popular files to be requested frequently by the MUs. The popularity distribution among the set $\boldsymbol{\mathcal F}$ is represented by $\boldsymbol {\mathcal P}=\{p_{1},p_{2},\cdots,p_{Q}\}$, where the MUs make independent requests of the $f$-th file, $f=1,\cdots,Q$, with the probability of $p_{f}$. Without any loss of generality, all these files have the same size of $M$ bits. We assume that $\mathcal B_0$ has a sufficiently large memory and hence accommodates the entire library of files, while the storage of each SBS is limited to $G$ files, where we have $G<Q$.

Without a loss of generality, we assume that ${Q}/{G}$ is an integer. The $Q$ files in $\boldsymbol{\mathcal F}$ are divided into $N={Q}/{G}$ file groups (FG), with each FG containing $G$ files. The $f$-th file, $\forall f\in\{(n-1)G+1,\cdots,nG\}$, is included in the $n$-th FG, $n\in\boldsymbol{\mathcal N}=\{1,\cdots,N\}$. We denote by $\mathcal F_n$ the $n$-th FG, and by $P_{\mathcal F_n}$ the probability that the MUs request a file in $\mathcal F_n$. Based on $\boldsymbol {\mathcal P}$, we have
\begin{equation}\label{equ:fil_blk_pop}
P_{\mathcal F_n}=\sum_{f=(n-1)G+1}^{nG}p_f.
\end{equation}
File caching is then carried out on the basis of FG, i.e., each SBS caches one of the $N$ FGs.
\section{Distributed File Downloading Relying on Caching}\label{sec:caching_protocol}
The caching-based distributed file downloading protocol consists of two stages. The first stage, or file placement stage, includes file content broadcasting and caching. In this stage, $\mathcal B_0$ broadcasts the FGs to the SBSs via the back-haul during off-peak periods. At the same time, the SBSs listen to the broadcasting from $\mathcal B_0$, and cache the FGs needed. The second stage, or file downloading stage, includes MU-SBS associations and file content transmissions. In this stage, each MU makes decisions as to which SBSs it should be associated with, and then starts to download files from the associated SBSs. When the requested files are not found in the adjacent SBSs, the MUs will turn to the MBS for these files.
\subsection{File Placement Matrix}
For assigning the $N$ FGs to the $K$ SBSs, we set up a file placement matrix $\boldsymbol \Lambda$ of size $K\times N$. The entry $\lambda_{k,n}\in\{0,1\}$ in $\boldsymbol \Lambda$ indicates whether $\mathcal F_n$ is cached by $\mathcal B_k$ or not. We have $\lambda_{k,n}=1$ if $\mathcal F_n$ is cached by $\mathcal B_k$, while $\lambda_{k,n}=0$ otherwise. The $k$-th row of $\boldsymbol \Lambda$ indicates which FG is cached by $\mathcal B_k$, and the $n$-th column indicates which BS caches $\mathcal F_n$. The number of the SBSs which cache $\mathcal F_n$ can be calculated as $\sum_{k\in\boldsymbol {\mathcal K}}\lambda_{k,n}$. Since each SBS caches one FG, we have $\sum_{n\in\boldsymbol {\mathcal N}}\lambda_{k,n}=1$.
\subsection{MU-SBS Association}
Denote by $\boldsymbol {\mathcal H}(j)$ the subscript set of the specific SBSs, which are capable of providing a sufficiently high SINR for the MU $\mathcal U_j$. The SBSs in $\boldsymbol {\mathcal H}(j)$ are the candidates for $\mathcal U_j$ to be potentially associated with. By setting an SINR threshold $\delta$, $\mathcal B_k$ will be included in $\boldsymbol {\mathcal H}(j)$ if and only if
\begin{equation}\label{equ:thres_SINR}
\frac{h_{k,j}^2d_{k,j}^{-\alpha}P_{k}}{\underset{{q\in{\boldsymbol{\mathcal K}}\backslash\{k\}}}{\sum}h_{q,j}^2d_{q,j}^{-\alpha}P_q+\sigma^2}\ge\delta.
\end{equation}

When requesting a file in $\mathcal F_n$, $\mathcal U_j$ first communicates with one of the SBSs in $\boldsymbol {\mathcal H}(j)$ which caches $\mathcal F_n$. It is possible that more than one SBS in $\boldsymbol {\mathcal H}(j)$ caches $\mathcal F_n$. In this case, $\mathcal U_j$ will associates with the optimal SBS, which imposes the minimum downloading delay.

It is clear that the downloading delay is inversely proportional to the downlink transmission rate. According to the file request assumption stipulated in the previous section, there is only a single MU connected to an SBS at each time. Thus, the maximum transmission rate from $\mathcal B_h$ to $\mathcal U_j$, $\forall h\in\boldsymbol{\mathcal H}(j)$, is the channel capacity between them, i.e., $C_{h,j}$. When $\mathcal U_j$ tries to download a file in $\mathcal F_n$, it follows the maximum-capacity association criterion. Hence, $\mathcal U_j$ associates with $B_{\hat h}$ such that
\begin{equation}\label{equ:association}
\hat h=\underset{{h\in{\boldsymbol {\mathcal H}(j)}}}{\arg\max}\{\lambda_{h,n}C_{h,j}\}.
\end{equation}

When none of the SBSs in $\boldsymbol {\mathcal H}(j)$ caches $\mathcal F_n$, i.e., we have $\lambda_{h,n}=0$, $\forall h\in{\boldsymbol {\mathcal H}(j)}$, $\mathcal U_j$ will associate with the MBS for the requested file.
\subsection{Optimization Problem Formulation}
We now optimize the matrix $\boldsymbol \Lambda$ for minimizing the average delay of downloading a file. Only when the optimal $\boldsymbol \Lambda$ has been determined will the file-placement stage commence, where the files are placed according this optimal matrix. Once the MU-SBS associations have been determined, we can optimize the matrix $\boldsymbol \Lambda$ for minimizing the average delay of downloading a file. First, given the channel coefficients and the specific location of $\mathcal U_j$, the delay of downloading a file in $\mathcal F_n$ by $\mathcal U_j$ can be calculated as
\begin{equation}\label{equ:Delay_U_j}
D_{j,n}=\begin{cases}
\frac{M}{\max_{h\in{\boldsymbol {\mathcal H}(j)}}\{\lambda_{h,n}C_{h,j}\}},&\exists \lambda_{h,n}\neq 0, \forall h\in{\boldsymbol {\mathcal H}(j)}\\
\frac{M}{C_{0}},&\text{otherwise}.
\end{cases}
\end{equation}
Based on the request probability of each FG, the delay for $\mathcal U_j$ to download a file from $\boldsymbol {\mathcal F}$ can be written as $D_{j}=\sum_{n\in\boldsymbol {\mathcal N}}P_{\mathcal F_n}D_{j,n}$. Thus, the average delay for each MU can be calculated as
\begin{equation}\label{equ:Overall_delay}
D=\frac1{J}\sum_{j\in\boldsymbol{\mathcal J}}D_{j}.
\end{equation}

By setting $D$ as the OF, let us hence formulate the delay optimization problem as follows:
\begin{equation}\label{equ:Optimzation_Express}
\begin{split}
\text{minimize} & \quad D\\
\text{s.t.} & \quad \sum_{n\in\boldsymbol {\mathcal N}}\lambda_{k,n}=1,\;\forall\; k\in\boldsymbol{\mathcal K},\\
&  \quad \boldsymbol \Lambda\in\{0,1\}^{K\times N}.
\end{split}
\end{equation}
The optimization problem in (\ref{equ:Optimzation_Express}) is an integer programming problem, which is NP-complete. In~\cite{Shuan:TIT13,Li:TMC13}, similar optimization problems have been solved by sub-optimal solutions, such as the classic greedy algorithm (GA). However, the existing solutions are typically based on centralized optimization. As we can see from (\ref{equ:Overall_delay}), a centralized minimization of $D$ at $\mathcal B_0$ requires the global CSI between $\boldsymbol {\mathcal B}$ and $\boldsymbol {\mathcal U}$, which is impractical. Hence, we will dispense with this assumption and optimize $\boldsymbol \Lambda$ in a distributed manner at a low complexity.
\begin{figure}
\centering
\includegraphics[width=\linewidth]{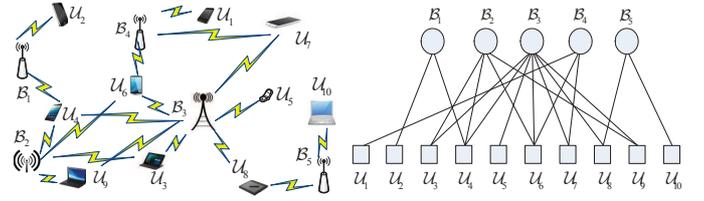}
\caption{Factor graph extracted from an HCN composed of $5$ SBSs and $10$ MUs. The edge between an SBS and an MU means that the SBS can provide a sufficiently high SINR for the MU. For instance, $\mathcal B_1$ can provide a sufficiently high SINR for $\mathcal U_2$ as well as $\mathcal U_4$. At the same time, $\mathcal U_3$ can receive a sufficiently high SINR from both $\mathcal B_2$ and $\mathcal B_3$.}\label{fig:Factor_G}
\end{figure}
\section{Distributed Belief Propagation Algorithm}\label{sec:cache_optimizations_BP}
In this section, we propose a distributed algorithm based on BP for solving the optimization problem of (\ref{equ:Optimzation_Express}) as follows: 1) We first develop a factor graph for describing the message passing in the BP algorithm. 2) Then we map the resultant factor graph to the network for the sake of facilitating the distributed BP optimization. 3) This solved by solving our optimization problem by proposing a distributed BP algorithm. 4) Finally, the proof of existence for a fixed point of convergence in the BP algorithm is presented.
\subsection{Factor Graph Model}
In our BP algorithm, the factor graph has to be first established based on the underlying network as a standard bipartite graphical representation of a mathematical relationship between the local delay functions and file allocation variables. Then the BP algorithm is implemented by iteratively passing messages between the local functions and their related variables. Our optimization problem is thus solved by the proposed BP algorithm based on the factor graph.

Based on the topology of the HCN, we develop a factor graph model $\boldsymbol {\mathcal G}=(\boldsymbol {\mathcal V},\boldsymbol {\mathcal E})$, where $\boldsymbol {\mathcal V}$ is the vertex set, and $\boldsymbol {\mathcal E}$ is the edge set. The vertex set $\boldsymbol {\mathcal V}$ consists of factor nodes and variable nodes. Each factor node is related to an MU and each variable node is related to an SBS. To simplify the notations, we denote by $j\in\boldsymbol{\mathcal J}$ the $j$-th factor node and denote by $k\in\boldsymbol{\mathcal K}$ the $k$-th variable node. Hence, the vertex set $\boldsymbol {\mathcal V}$ is composed of $\boldsymbol {\mathcal J}$ and $\boldsymbol {\mathcal K}$, i.e., $\boldsymbol {\mathcal V}=\{\boldsymbol {\mathcal J},\boldsymbol {\mathcal K}\}$.

As mentioned in the previous section, $\mathcal B_k$ will be a candidate for $\mathcal U_j$ to potentially associate with, but only if the received SINR at $\mathcal U_j$ from $\mathcal B_k$ is no less than the threshold $\delta$. Correspondingly, in our factor graph, an edge in the edge set $\boldsymbol {\mathcal E}$ connecting $\mathcal U_j$ and $\mathcal B_k$, denoted by $(j,k)$, exists if the received SINR at $\mathcal U_j$ from $\mathcal B_k$ is no less than $\delta$. The node $k$ is named as a neighboring node of $j$, if there is an edge $(j,k)$. Actually, $\boldsymbol{\mathcal H}(j)$ defined previously represents the set of the neighboring nodes of the factor node $j$. Furthermore, denote by $\boldsymbol{\mathcal H}(k)$ the set of neighboring node for the variable node $k$. Fig.~\ref{fig:Factor_G} illustrates a factor graph extracted from an HCN with $5$ SBSs and $10$ MUs. Take $\mathcal B_1$ in the factor graph for example. The edges exist between $\mathcal B_1$ and $\mathcal U_2$ as well as $\mathcal U_4$, which means that $\mathcal B_1$ can provide a sufficient large SINR for both $\mathcal U_2$ and $\mathcal U_4$.

The distributed BP algorithm is based on the factor graph $\boldsymbol {\mathcal G}$. The factor nodes in $\boldsymbol{\mathcal J}$ represent the local utility functions generated from the decomposition results of the global utility function, which will be discussed later in this subsection. The variable nodes in $\boldsymbol {\mathcal K}$ represent the variables to be optimized, i.e., the entries of $\boldsymbol \Lambda$. The factor nodes and variable nodes are connected by edges in $\boldsymbol{\mathcal E}$, indicating the message flows in the BP algorithm. That is, messages are only passing between a node and its neighbors. We now illustrate the optimization problem on the factor graph.
\subsubsection{Factor Nodes}
According to Eq. (\ref{equ:Optimzation_Express}), the OF can be decomposed into $J$ local contributions as $D_1,\cdots, D_J$. These local contributions are calculated based on Eq.~(\ref{equ:Delay_U_j}). Since the BP algorithm solves maximization problems, we define a series of utility functions as $F\triangleq -D$ and $F_j\triangleq -D_j$. Then our optimization problem can be rewritten as
\begin{equation}\label{equ:inverse_problem}
\max_{\boldsymbol\Lambda}F({\boldsymbol\Lambda}), \quad F= \frac1{J}\sum_{j\in\mathcal J}F_j.
\end{equation}
We use the $j$-th factor node to represent the $j$-th local utility function $F_j$, which is related to $\mathcal U_j$. Hence, the maximization of $F$ can be achieved by maximizing $F_j$ at $\mathcal U_j$, $\forall j\in\boldsymbol {\mathcal J}$.
\subsubsection{Variable Nodes}
Each variable node is related to an SBS. Here, we use the $k$-th variable node to represent the $k$-th row of $\boldsymbol \Lambda$, denoted by $\boldsymbol\lambda_k$, which is related to $\mathcal B_k$. The location of `1' in $\boldsymbol\lambda_k$ indicates which specific FG is stored by $\mathcal B_k$. Note that the first constraint in (\ref{equ:Optimzation_Express}) means that each SBS only stores a single FG. Given this constraint, $\boldsymbol\lambda_k$ has $N$ possible values according to $N$ different locations of `1'. We denote by $\boldsymbol\lambda_k^{[1]}$, $\cdots$, $\boldsymbol\lambda_k^{[N]}$ the $N$ values of $\boldsymbol\lambda_k$. When we have $\boldsymbol\lambda_k=\boldsymbol\lambda_k^{[n]}$, this implies that the FG $\mathcal F_n$ is stored by $\mathcal B_k$. Take $N=2$ for example, where $\boldsymbol\lambda_k=\boldsymbol\lambda_k^{[1]}=[1~0]$ indicates that the FG $\mathcal F_1$ is stored in the SBS $\mathcal B_k$, while $\boldsymbol\lambda_k=\boldsymbol\lambda_k^{[2]}=[0~1]$ indicates that $\mathcal F_2$ is stored in $\mathcal B_k$. The variables $\boldsymbol\lambda_k$, $k=1,\cdots,K$, are the parameters to be optimized for maximizing $F$ in (\ref{equ:inverse_problem}). For simplicity, we use the matrix $\boldsymbol\Lambda$ to represent the set of the variables $\boldsymbol\lambda_k$ in the factor graph.
\subsection{Distributed Belief Propagation}
In standard BP, the variables are optimized by estimating their marginal probability distributions~\cite{Rangan:JSAC12}. Note that the utility function $F$ is a function of the file placement matrix $\boldsymbol \Lambda$. We define the probability mass function (PMF) $p(\boldsymbol \Lambda)$ of $\boldsymbol \Lambda$ based on the utility function $F(\boldsymbol \Lambda)$ as
\begin{equation}\label{equ:pdf_utility}
p(\boldsymbol \Lambda)\triangleq \frac1{Z}\exp\left(\mu F(\boldsymbol \Lambda)\right),
\end{equation}
where $\mu$ is a positive number and $Z$ is the normalization factor. According to~\cite{Rangan:JSAC12}, the result of large deviations shows that when $\mu\rightarrow\infty$, $p(\boldsymbol \Lambda)$ concentrates around the maxima of $F(\boldsymbol\Lambda)$, i.e., $\lim_{\mu\rightarrow\infty} \mathbb E(\boldsymbol \Lambda)=\underset{\boldsymbol \Lambda}{\arg\max}F(\boldsymbol\Lambda)$, where $\mathbb E(\boldsymbol \Lambda)$ is the expectation of $\boldsymbol \Lambda$. Once we obtain $\mathbb E(\boldsymbol \Lambda)$, we can have a good estimate of the specific $\boldsymbol \Lambda$ which maximizes $F(\boldsymbol\Lambda)$.

In our distributed BP, the maximization of $F$ can be decomposed into $J$ maximization operations on $F_j$ at $\mathcal U_j$, $j=1,\cdots, J$. Correspondingly, the estimation of $\boldsymbol \Lambda$ is decomposed into $J$ estimations of its subsets $\boldsymbol \Lambda_j$ at $\mathcal U_j$, where $\boldsymbol \Lambda_j=\{\boldsymbol \lambda_h,\forall h\in\boldsymbol {\mathcal H}(j)\}$. The PMF of $\boldsymbol \Lambda_j$ is written as $p_j(\boldsymbol \Lambda_j)=\frac1{Z_j}\exp\left(\mu F_j(\boldsymbol\Lambda_j)\right)$, where $Z_j$ is the normalization factor. Since all the variables are independent, the estimation of $\boldsymbol \Lambda_j$ at $\mathcal U_j$ can be further decomposed into the estimation of each individual $\boldsymbol \lambda_h$ via calculating its PMF $p_j(\boldsymbol \lambda_h)$, which is the marginal PMF of $p_j(\boldsymbol\Lambda_j)$ with respect to the variable $\boldsymbol{\lambda}_{h}$. Hence we have $p_j(\boldsymbol{\lambda}_h)=\mathbb E_{\sim{\boldsymbol{\lambda}_h}}(p_j(\boldsymbol\Lambda_j))$, where $\mathbb E_{\sim{\boldsymbol{\lambda}_h}}(\cdot)$ represents the expectation over the elements in $\boldsymbol\Lambda_j$, except for $\boldsymbol{\lambda}_h$. The PMF $p_j(\boldsymbol{\lambda}_h)$ is viewed as the message, which is iteratively updated between $\mathcal U_j$ and $\mathcal B_{h}$, $\forall h\in\boldsymbol {\mathcal H}(j)$. The PMF $p_j(\boldsymbol{\lambda}_h)$ consists of $N$ probabilities estimated by $\mathcal U_j$, i.e., $\Pr(\boldsymbol{\lambda}_h=\boldsymbol{\lambda}_h^{[1]}),\cdots,\Pr(\boldsymbol{\lambda}_h=\boldsymbol{\lambda}_h^{[N]})$, where $\Pr(\boldsymbol{\lambda}_h=\boldsymbol{\lambda}_h^{[n]})$ represents the probability that $\mathcal F_n$ is stored by $\mathcal B_h$.

Without a loss of generality, we assume that the edge $(j,k)$ does exist in the factor graph. We represent the iteration index by $t$ and denote by $p_{k\rightarrow j}^{(t)}(\boldsymbol\lambda_k)$ and $p_{j\rightarrow k}^{(t)}(\boldsymbol\lambda_k)$ the belief messages emanated from $\mathcal B_k$ to $\mathcal U_j$ and from $\mathcal U_j$ to $\mathcal B_k$ during the $t$-th iteration, respectively. The steps describing the distributed BP are as follows.
\subsubsection{Initialization}
At the variable nodes, set $t=1$ and let $p_{k\rightarrow j}^{(1)}(\boldsymbol\lambda_k)$ to be the initial distribution of $\boldsymbol\lambda_k$, e.g., the \emph{a priori} popularity distribution $\boldsymbol{\mathcal P}$.
\subsubsection{Variable Node Update}
During the $t$-th iteration, each SBS $\mathcal B_k$ updates the message $p_{k\rightarrow j}^{(t)}(\boldsymbol\lambda_k)$ to be sent to $\mathcal U_j$ based on the messages gleaned from $\mathcal B_k$'s neighboring MUs other than $\mathcal U_j$ in the previous iteration. This includes the calculations of $N$ probabilities. Given $\boldsymbol{\lambda}_k=\boldsymbol{\lambda}_k^{[n]}$, $\forall n\in\boldsymbol{\mathcal N}$, we have
\begin{equation}\label{equ:variable_update}
p_{k\rightarrow j}^{(t)}(\boldsymbol\lambda_k^{[n]})=\frac1{Z_k}\prod_{\hbar\in{\boldsymbol {\mathcal H}(k)}\backslash\{j\}}p_{\hbar\rightarrow k}^{(t-1)}(\boldsymbol\lambda_k^{[n]}),
\end{equation}
where $Z_k$ is the normalization factor so that we have $\sum_{n\in{\boldsymbol {\mathcal N}}}p_{k\rightarrow j}^{(t)}(\boldsymbol\lambda_k^{[n]})=1$.
\subsubsection{Factor Node Update}
In the $t$-th iteration, $\mathcal U_j$ updates the $N$ probabilities of the message $p_{j\rightarrow k}^{(t)}(\boldsymbol\lambda_k)$ to be sent to $\mathcal B_k$, which is based on the messages received from $\mathcal U_j$'s neighboring SBSs, except for $\mathcal B_k$. The messages updated at the factor nodes are calculated according to the marginal PMF. Given $\boldsymbol{\lambda}_k=\boldsymbol{\lambda}_k^{[n]}$, $\forall n\in\boldsymbol{\mathcal N}$, we have
\begin{equation}\label{equ:factor_update}
\begin{split}
&p_{j\rightarrow k}^{(t)}(\boldsymbol\lambda_k^{[n]})\\&=\mathbb E_{\sim\boldsymbol\lambda_k}\left(\exp\left(\mu F_j\left(\boldsymbol {\lambda}_k^{[n]},\{\boldsymbol \lambda_h,\forall h\in\boldsymbol{\mathcal H}(j)\backslash\{k\}\}\right)\right)\right)\\&
=\sum_{h\in\boldsymbol{\mathcal H}(j)\backslash\{k\}}\sum_{\boldsymbol\lambda_h=\boldsymbol\lambda_{h}^{[1]}}^{\boldsymbol\lambda_{h}^{[N]}}\bigg(\prod_{q\in\boldsymbol{\mathcal H}(j)\backslash\{k\}}
p_{q\rightarrow j}^{(t)}(\boldsymbol\lambda_q)\cdot\\&\qquad\exp\left(\mu F_j\left(\boldsymbol {\lambda}_k^{[n]},\left\{\boldsymbol \lambda_h,\forall h\in\boldsymbol{\mathcal H}(j)\backslash\{k\}\right\}\right)\right)\bigg).
\end{split}
\end{equation}
\subsubsection{Final Solution}
Let us assume that there are $t=T$ iterations in the distributed BP algorithm. After $T$ iterations, the probability that $\mathcal F_n$ is stored by $\mathcal B_k$ can be obtained by
\begin{equation}\label{equ:final_solution}
\Pr(\boldsymbol\lambda_k=\boldsymbol\lambda_k^{[n]})=\frac1{Z_k}\prod_{\hbar\in{\boldsymbol {\mathcal H}(k)}}p_{\hbar\rightarrow k}^{(T)}(\boldsymbol\lambda_k^{[n]}).
\end{equation}
Based on (\ref{equ:final_solution}), the decision as to which file should be stored by $\mathcal B_k$ can be made by choosing the specific file that has the maximum \emph{a posteriori} probability $\Pr(\boldsymbol\lambda_k=\boldsymbol\lambda_k^{[n]})$, $\forall n\in\boldsymbol{\mathcal N}$.
\subsection{Convergence to a Fixed Point}
Let us now investigate the existence of a fixed point of convergence in our distributed BP algorithm. The essence of the distributed BP algorithm is to keep updating the PMF $p_j(\boldsymbol{\lambda}_k)$ before reaching its final estimate. Based on (\ref{equ:variable_update}) and (\ref{equ:factor_update}), the evolution of $p_j(\boldsymbol{\lambda}_k)$ during the $t$-th iteration can be obtained from the PMFs in the $(t-1)$-th iteration as
\begin{equation}\label{equ:PMF_evolution}
\begin{split}
p_{k\rightarrow j}^{(t)}(\boldsymbol\lambda_k)=&\frac1{Z_k}\prod_{\hbar\in{\boldsymbol {\mathcal H}(k)}\backslash\{j\}}\sum_{h\in\boldsymbol{\mathcal H}(\hbar)\backslash\{k\}}\sum_{\boldsymbol\lambda_h=\boldsymbol\lambda_{h}^{[1]}}^{\boldsymbol\lambda_{h}^{[N]}}\\&\left(\exp(\mu F_\hbar(\boldsymbol{\Lambda}_\hbar))\cdot\prod_{q\in\boldsymbol{\mathcal H}(\hbar)\backslash\{k\}}
p_{q\rightarrow \hbar}^{(t-1)}(\boldsymbol\lambda_q)\right).
\end{split}
\end{equation}
We view the PMF $p_{k\rightarrow j}^{(t)}(\boldsymbol\lambda_k)$ as a probability vector of length $N$. We define the probability vector set $\boldsymbol{\mathcal M}^{(t)}\triangleq\left\{p_{k\rightarrow j}^{(t)}(\boldsymbol\lambda_k)\right\}$ for all $k\in\boldsymbol{\mathcal K}$ as well as $j\in\boldsymbol{\mathcal J}$, and define the message mapping function $\boldsymbol\Gamma:\mathbb{R}^{N\times KJ}\rightarrow\mathbb{R}^{N\times KJ}$ based on (\ref{equ:PMF_evolution}) so that $\boldsymbol{\mathcal M}^{(t)}=\boldsymbol\Gamma(\boldsymbol{\mathcal M}^{(t-1)})$. Then we have the following lemma.
\begin{lemma}\label{lem:continous_mapping}
The message mapping function $\boldsymbol\Gamma$ is a continuous mapping.

\emph{Proof:} Please refer to Appendix~\ref{app:continous_mapping}.
\end{lemma}

Given Lemma \ref{lem:continous_mapping}, we have the following theorem.
\begin{theorem}\label{theo:fixed_point}
A fixed point of convergence exists for the proposed distributed BP algorithm.

\emph{Proof:} Please refer to Appendix~\ref{app:fixed_point}.
\end{theorem}

The question of convergence to the fixed point is, unfortunately, not well understood in general~\cite{Moallem:INFO11}. Generally, if the factor graph contains no cycles, the belief propagation can be shown to converge to a fixed solution point in a finite number of iterations. The performance, including the optimality and the convergence rate, of the BP crucially depends on the choice of the objective function, as well as the scale, the sparsity and the number of cycles in the underlying factor graph. As such, the theoretical analysis of the BP algorithm's optimality and convergence rate remains an open challenge.
\section{A Heuristic BP with Reduced Complexity}\label{sec:heuristic_BP}
In the context of the BP algorithm, the message $p_j(\boldsymbol \lambda_k)$ exchanged between $\mathcal U_j$ and $\mathcal B_k$ in each iteration, includes $N$ probability values, which are real numbers. Hence, the communication overhead of the message passing is relatively high. Hence, we propose a heuristic BP (HBP) algorithm for reducing the communication overhead imposed. The rationale behind the term ``heuristic BP'' is that we still follow the classic concept of belief propagation, but use a different format of the beliefs from the conventional one.

Assuming that the edge $(j,k)$ exists, in the $t$-th iteration of the HBP, instead of forwarding the $N$ probabilities stored in $p^{(t)}_{j\rightarrow k}(\boldsymbol \lambda_k)$ to $\mathcal B_k$, $\mathcal U_j$ randomly selects an FG according to these $N$ probabilities. Then the integer index $n^{(t)}_{j\rightarrow k}$ of the FG selected will be forwarded to the SBS $\mathcal B_k$.

At the SBS side, the SBS $\mathcal B_k$ receives $|\boldsymbol{\mathcal H}(k)|$ integers, i.e., $n^{(t)}_{\hbar\rightarrow k}$, $\forall \hbar\in\boldsymbol{\mathcal H}(k)$, from its neighboring MUs, where $|\cdot|$ denotes the cardinality of a set. Based on $n^{(t)}_{\hbar\rightarrow k}$, the SBS $\mathcal B_k$ infers the number of those MUs, which indicate that $\mathcal F_n$ should be stored in the SBS $\mathcal B_k$, for $n=1,\cdots, N$. Let us assume now that in the $t$-th iteration, there are $J_{k,n}^{(t)}$ MUs specifically indicating that $\mathcal F_n$ should be stored in $\mathcal B_k$, where we have $\sum_{n\in\boldsymbol{\mathcal N}} J_{k,n}^{(t)}=|\boldsymbol{\mathcal H}(k)|$. We can view $\frac{J_{k,n}^{(t)}}{|\boldsymbol{\mathcal H}(k)|}$ as the probability that the specific FG $\mathcal F_n$ is stored by the SBS $\mathcal B_k$.

In this case, the probability $p_{k\rightarrow j}^{(t)}(\boldsymbol\lambda_k^{[n]})$ in (\ref{equ:variable_update}) will be recalculated as
\begin{equation}\label{equ:variable_update_1}
p_{k\rightarrow j}^{(t)}(\boldsymbol\lambda_k^{[n]})=\begin{cases}
\frac{J_{k,n}^{(t-1)}-1}{|\boldsymbol{\mathcal H}(k)|-1},&\text{if}\;\; n=n^{(t-1)}_{j\rightarrow k},\\
\frac{J_{k,n}^{(t-1)}}{|\boldsymbol{\mathcal H}(k)|-1},&\text{if}\;\; n\neq n^{(t-1)}_{j\rightarrow k}.
\end{cases}
\end{equation}
Note that in (\ref{equ:variable_update_1}), the information $n^{(t-1)}_{j\rightarrow k}$ transmitted from the MU $\mathcal U_j$ to the SBS $\mathcal B_k$ is excluded when calculating $p_{k\rightarrow j}^{(t)}(\boldsymbol\lambda_k^{[n]})$, for the sake of ensuring that only uncorrelated information is exchanged throughout the HBP.

At the MU side, it is clear that the MU $\mathcal U_j$ has to obtain $p_{k\rightarrow j}^{(t)}(\boldsymbol\lambda_k^{[n]})$ for the sake of updating the output information. However, there is no need for the SBS $\mathcal B_k$ to transmit the $N$ probabilities $p_{k\rightarrow j}^{(t)}(\boldsymbol\lambda_k^{[n]})$ to each of its neighboring MUs. Alternatively, $\mathcal B_k$ broadcasts the $N$ integers, $J_{k,1}^{(t)},\cdots,J_{k,N}^{(t)}$ to the neighboring MUs for reducing the transmission overhead. After receiving the $N$ integers from the SBS $\mathcal B_k$, the MU $\mathcal U_j$ calculates $p_{k\rightarrow j}^{(t)}(\boldsymbol\lambda_k^{[n]})$ in (\ref{equ:variable_update_1}).

Based on the above discussions, the HBP algorithm can be summarized as follows.
\subsubsection{Initialization}
At the variable nodes, we set $t=1$. The SBS $\mathcal B_k$ randomly generates $|\boldsymbol{\mathcal H}(k)|$ independent integers, $n_1,\cdots, n_{|\boldsymbol{\mathcal H}(k)|}$, according to the popularity distribution $\boldsymbol{\mathcal P}$. These integers are viewed as the indexes of the FGs. We then set $J_{n,k}^{(1)}$ to be the number of the integers that are equal to $n$.
\subsubsection{Variable Node Update}
In the $t$-th iteration, $\mathcal B_k$ updates and broadcasts the $N$ integers $J_{n,k}^{(t)}$, for $n=1,\cdots, N$, to the neighboring MUs. The resulting calculations performed on these $N$ integers $J_{n,k}^{(t)}$ are based on the integers $n^{(t-1)}_{\hbar\rightarrow k}$, $\forall \hbar\in\boldsymbol{\mathcal H}(k)$, received from the neighboring MUs during the last iteration. Specifically, the $n$-th integer $J_{n,k}^{(t)}$ is obtained by counting the number of $n^{(t-1)}_{\hbar\rightarrow k}$ that are equal to $n$.
\subsubsection{Factor Node Update}
The MU $\mathcal U_j$ first calculates the probabilities $p_{h\rightarrow j}^{(t)}(\boldsymbol\lambda_k^{[n]})$, $\forall h\in\boldsymbol {\mathcal H}(j)$ according to Eq. (\ref{equ:variable_update_1}) based on the integers gleaned from the SBS $\mathcal B_h$. Then based on $p_{h\rightarrow j}^{(t)}(\boldsymbol\lambda_k^{[n]})$, $\forall h\in\boldsymbol {\mathcal H}(j)\backslash\{k\}$, $\mathcal U_j$ calculates $p_{j\rightarrow k}^{(t)}(\boldsymbol\lambda_k^{[n]})$ according to Eq. (\ref{equ:factor_update}). After obtaining the $N$ probabilities $p_{j\rightarrow k}^{(t)}(\boldsymbol\lambda_k^{[n]})$, $n=1,\cdots, N$, $\mathcal U_j$ randomly chooses an FG according to these $N$ probabilities and sends the index $n_{j\rightarrow k}^{(t)}$ of the FG to the SBS $\mathcal B_k$.
\subsubsection{Final Solution}
After $T$ iterations, the SBS $\mathcal B_k$ makes the decision that the FG $\mathcal F_{\hat n}$ should be stored for ensuring that
\begin{equation}\label{equ:final_nhat}
\hat n=\underset{n\in\boldsymbol{\mathcal N}}{\arg\max}{J_{k,n}^{(T)}}.
\end{equation}

The overhead of the HBP is significantly lower than that of the original BP introduced in the previous section. From a communication complexity perspective, in each iteration of the HBP, an SBS ${\mathcal B}_k$ broadcasts $N$ integers, while an MU $ {\mathcal U}_j$ transmits $|\boldsymbol{\mathcal H}(j)|$ integers. On the other hand, in the original BP, ${\mathcal B}_k$ transmits $N|\boldsymbol{\mathcal H}(k)|$ real numbers, while ${\mathcal U}_j$ transmits $N|\boldsymbol{\mathcal H}(j)|$ real numbers for each iteration. From a computational complexity perspective, in a single iteration of the HBP, the computational complexity is on the order of $O(N)$ at the SBS ${\mathcal B}_k$, and $O(|{\mathcal H}(j)|N^{|\mathcal H(j)|})$ at the MU ${\mathcal U}_j$. On the other hand, in the original BP, the computational complexity is $O(N|\mathcal H(k)|^2)$ at ${\mathcal B}_k$, and $O(|{\mathcal H}(j)|N^{|\mathcal H(j)|})$ at ${\mathcal U}_j$ for each iteration.
\section{Performance Analysis based on Stochastic Geometry}\label{sec:stochastic}
In this section, we analyze both the average degree distribution of the factor graph and the average downloading performance based on stochastic geometry theory. We model the distribution of the MUs as a PPP $\Phi_{U}$ having the intensity of $\lambda_U$, and that of the SBSs as an independent PPP $\Phi_{B}$ with the intensity $\lambda_B$~\cite{Dhillon:JSAC12,Han-Shin:TWC12}. For simplicity, we assume that all the SBSs have the same transmission power $P$. In the following, both the degree distribution and the downloading performance are averaged over both the channels' fading coefficients and over the PPP distributions of the nodes.
\subsection{Average Degree Distributions of the Factor Graph}
Let us now investigate the degree distribution of the factor graph averaged over PPP. Note that the degree of a factor node $j$ is defined as the number of its neighboring variable nodes, given by the cardinality $|\boldsymbol {\mathcal H}(j)|$, while the degree of a variable node $k$ is defined as the number of its neighboring factor nodes, i.e., $|\boldsymbol {\mathcal H}(k)|$. Then we have the following theorem.
\begin{theorem}\label{theo:average}
The factor nodes in the factor graph have the average degree
\begin{equation}\label{Eq:average_MU}
\zeta_{U}=2\pi\lambda_{B}Z\left(\lambda_B, P, \alpha, \delta \right),
\end{equation}
and the variable nodes have the average degree
\begin{equation}\label{Eq:average_BS}
\zeta_{B}=2\pi\lambda_{U}Z\left(\lambda_{B}, P, \alpha, \delta \right),
\end{equation}
where we have
\begin{multline}\label{equ:z_func}
Z\left(\lambda_{B}, P, \alpha, \delta\right)=\\\int^{\infty}_{0}\exp\left\{-\frac{2\lambda_{B}\pi}{\alpha}\delta ^{\frac{2}{\alpha}}B\left(\frac{2}{\alpha},1-\frac{2}{\alpha}\right)r^2-\frac{\delta \sigma^{2}}{P}r^{\alpha}\right\}r\text{d}r
\end{multline}
and the Beta function $B(x,y)=\int^{1}_{0}t^{x-1}(1-t)^{y-1}\text{d}t$.

\emph{Proof:} Please refer to Appendix~\ref{app:average}.
\end{theorem}

When neglecting the noise, we have the following corollary based on \emph{Theorem~\ref{theo:average}}.
\begin{corollary}\label{cor:alpha}
When neglecting the noise, $Z\left(\lambda_{B}, P, \alpha, \delta\right)$ in (\ref{equ:z_func}) can be rewritten as
\begin{equation}\label{equ:approx_z}
Z\left(\lambda_{B}, P, \alpha, \delta\right)=\frac{\alpha}{4\pi\lambda_B B\left(\frac{2}{\alpha},1-\frac{2}{\alpha}\right) \delta^{\frac{2}{\alpha}}}.
\end{equation}
Then we can simplify the average degree of the factor nodes in Eq. (\ref{Eq:average_MU}) to
\begin{equation}\label{equ:approx_du}
\zeta_{U}=\frac{\alpha}{2 \delta^{\frac{2}{\alpha}} B\left(\frac{2}{\alpha}, 1-\frac{2}{\alpha}\right)},
\end{equation}
and the average degree of the variable nodes in Eq. (\ref{Eq:average_BS}) to
\begin{equation}\label{equ:approx_dl}
\begin{split}
\zeta_{B}=\frac{\lambda_U\alpha}{2 \lambda_B\delta^{\frac{2}{\alpha}} B\left(\frac{2}{\alpha}, 1-\frac{2}{\alpha}\right)}.
\end{split}
\end{equation}

\emph{Proof:} Please refer to Appendix~\ref{app:MU_nolambda}.
\end{corollary}

Equations (\ref{equ:approx_du}) and (\ref{equ:approx_dl}) can be seen as approximations of (\ref{Eq:average_MU}) and (\ref{Eq:average_BS}), respectively, when the effects of the noise are neglected. These approximations are significantly accurate for the HCN, since the interference effects are dominant due to the dense deployments of the SBSs.

From (\ref{equ:approx_du}), we can see that $\zeta_U$ is only related to $\delta$ and $\alpha$, but is independent of $\lambda_U$, $P$ and $\lambda_B$. In other words, the factor node degree has no relation with the intensities of the MUs and SBSs or with the power of the SBSs. The intuitive reason is that although increasing both the PPP intensities and the power of the SBSs can increase the total signal power, the interference also increases at the same time, which keeps the degree $\zeta_U$ of the factor nodes constant. Similarly, observe from (\ref{equ:approx_dl}) that $\zeta_B$ is independent of the power $P$, i.e., increasing the transmission power of the SBSs will not influence the average degree distribution of the factor graph.

\begin{remark}\label{rmk:alpha4}
We observe that $B\left(\frac{2}{\alpha}, 1-\frac{2}{\alpha}\right)=\pi$ when $\alpha=4$. Thus, we have closed-form expressions for $\zeta_U$ and $\zeta_B$ in (\ref{equ:approx_du}) and (\ref{equ:approx_dl}), respectively, when $\alpha=4$.
\end{remark}
\subsection{Downloading Performance of Random Caching}
Since the performance of BP based caching remains difficult for mathematical analysis in closed form, we propose a random caching scheme and analyze its performance based on stochastic geometry theory. The random caching is realized by randomly picking out ${\Omega}_{\mathcal F_n}\cdot K$ ($0\le\Omega_{\mathcal F_n}\le1$) SBSs from the entire set of $K$ SBSs for caching the FG $\mathcal F_n$.

To evaluate the downloading performance, we first define an outage $\mathcal Q_n$ as the event of an MU's failing to find the FG $\mathcal F_n$ in its neighboring SBSs. The following theorem states an upper bound of the OP of $\mathcal Q_n$. As mentioned before, since the interference is the dominant factor predetermining the network performance, we ignore the noise effects in the following performance analysis to simplify our derivations.
\begin{theorem}\label{theo:outage_prob}
The OP for downloading a file in $\mathcal F_n$ can be upper-bounded by
\begin{equation}\label{equ:Qn}
\Pr(\mathcal{Q}_n)\le\frac{C(\delta, \alpha)(1-\Omega_{\mathcal F_n})+\emph{A}(\delta,\alpha)\Omega_{\mathcal F_n}}{C(\delta, \alpha)(1-\Omega_{\mathcal F_n})+\emph{A}(\delta,\alpha)\Omega_{\mathcal F_n}+\Omega_{\mathcal F_n}},
\end{equation}
where we have $C(\delta,\alpha)\triangleq\frac{2}{\alpha}\delta^{\frac{2}{\alpha}}B\left(\frac{2}{\alpha},1-\frac{2}{\alpha}\right)$, $\emph{A}(\delta,\alpha)\triangleq\frac{2\delta}{\alpha-2}\; {{}_2}F_{1}\left(1,1-\frac{2}{\alpha};2-\frac{2}{\alpha};-\delta\right)$, and ${{}_2}F_{1}$ represents the hypergeometric function.

\emph{Proof:} Please refer to Appendix~\ref{app:outage_prob}.
\end{theorem}

When the path-loss exponent $\alpha=4$, we have $C(\delta,4)=\frac{\sqrt\delta}{2}\pi$ and $A(\delta,4)=\delta{~}_2F_1(1,\frac1{2};\frac{3}{2},-\delta)$. It becomes clear from (\ref{equ:Qn}) that $\Pr(\mathcal{Q}_n)$ is only related to $\delta$ and $\Omega_{\mathcal F_n}$, where a higher $\delta$ leads to a higher $\Pr(\mathcal{Q}_n)$. This is because a larger $\delta$ will reduce the number of possibly eligible serving SBSs, resulting in an increase of OP. We can see that a higher $\Omega_{\mathcal F_n}$ leads to a lower $\Pr(\mathcal{Q}_n)$.

Let us define the averaged OP $\mathcal Q$ over all the files. Based on the file popularity, the OP of $\mathcal Q$ can be upper-bounded by
\begin{multline}\label{equ:aver_outage}
\Pr(\mathcal Q)=\sum_{n\in{{\boldsymbol {\mathcal N}}}}P_{\mathcal F_n}\Pr(\mathcal Q_n)\\\le\sum_{n\in{{\boldsymbol {\mathcal N}}}}\frac{P_{\mathcal F_n}(C(\delta, \alpha)(1-\Omega_{\mathcal F_n})+\emph{A}(\delta,\alpha)\Omega_{\mathcal F_n})}{C(\delta, \alpha)(1-\Omega_{\mathcal F_n})+\emph{A}(\delta,\alpha)\Omega_{\mathcal F_n}+\Omega_{\mathcal F_n}}.
\end{multline}
The average delay $\bar{D}$ of each MU can be obtained based on the average OP, i.e.,
\begin{equation}\label{equ:downloading_ave}
\bar{D}=(1-\Pr(\mathcal{Q}))\bar D_{s}+\Pr(\mathcal{Q})\frac{M}{C_0},
\end{equation}
where $\bar D_{s}$ is the average delay of downloading from the SBSs. The delay $\bar D$ can be seen as the average value of $D$ in Eq. (\ref{equ:Overall_delay}) over both the PPP and the channel fading. Note that $\bar D_{s}$ is usually challenging to calculate and does not have a closed form in the PPP analysis.

Next, we optimize $\Omega_{\mathcal F_n}$ for improving the downloading performance. Since we do not have a closed-form expression for $\bar{D}$, we minimize the upper bound of $\Pr(\mathcal{Q})$ in (\ref{equ:aver_outage}), i.e.,
\begin{equation}\label{equ:min_outage}
\begin{split}
\max_{\{\Omega_{\mathcal F_n}\}}& \sum_{n\in\boldsymbol {\mathcal N}}\frac{P_{\mathcal F_n} \Omega_{\mathcal F_n}}{\Omega_{\mathcal F_n}\left(\emph{A}(\delta,\alpha)-\emph{C}(\delta,\alpha)+1\right)+\emph{C}(\delta,\alpha)},\\
\text{s.t.} &\sum_{n\in\boldsymbol{\mathcal N}}\Omega_{\mathcal F_n}=1, \\
&\Omega_{\mathcal F_n}\geq 0.
\end{split}
\end{equation}
By relying on the classic Lagrangian multiplier, we arrive at the optimal solution as
\begin{equation}\label{equ:optimal_omega}
\Omega^\star_{\mathcal F_n}=\max\left\{\frac{\sqrt{\frac{P_{\mathcal F_n}}{\xi}}-\emph{C}(\delta,\alpha)}{\emph{A}(\delta,\alpha)-\emph{C}(\delta,\alpha)+1},0\right\},
\end{equation}
where $\xi=\frac{\left(\sum^{n^{*}}_{q=1}\sqrt{P_{\mathcal F_q}}\right)^2}{\left(n^{*}\emph{C}(\delta,\alpha_s)+\emph{A}(\delta,\alpha_s)-\emph{C}(\delta,\alpha_s)+1\right)^2}$, and $n^{*}$ satisfies the constraint that $\Omega_{\mathcal F_n}\geq 0$.

\begin{figure}
\centering
\includegraphics[width=\linewidth]{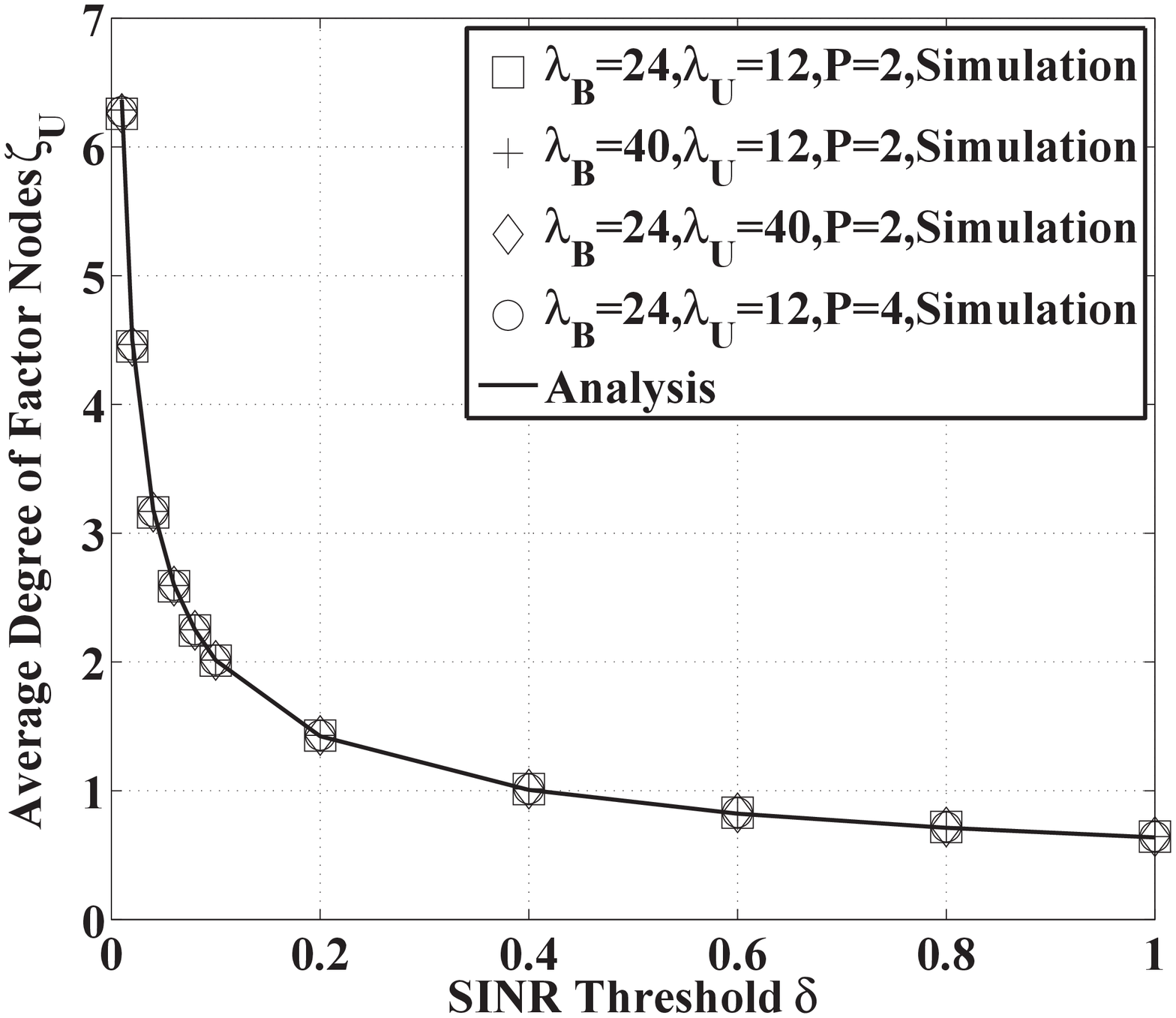}
\caption{Average degree of factor nodes $\zeta_U$ vs. $\delta$ for different SBS and MU intensities of $\lambda_B$ and $\lambda_U$, and for transmit powers of $P=2$ and 4.}\label{fig:Zeta_U}
\end{figure}
\begin{figure}
\centering
\includegraphics[width=\linewidth]{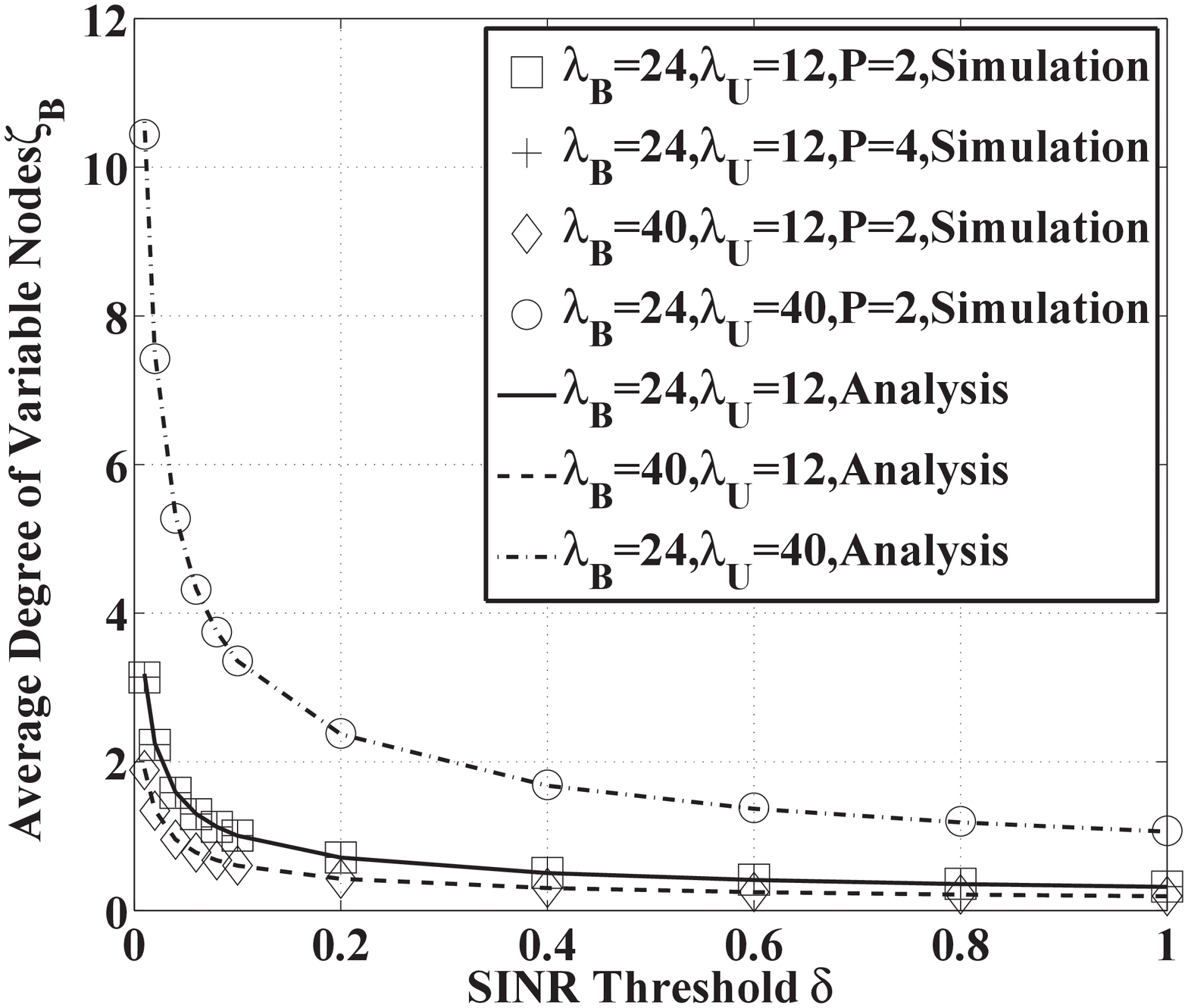}
\caption{Average degree of variable nodes $\zeta_B$ vs. $\delta$ for different SBS and MU intensities of $\lambda_B$ and $\lambda_U$, and for transmit powers of $P=2$ and 4.}\label{fig:Zeta_B}
\end{figure}
\begin{figure}[t]
\centering
\includegraphics[width=\linewidth]{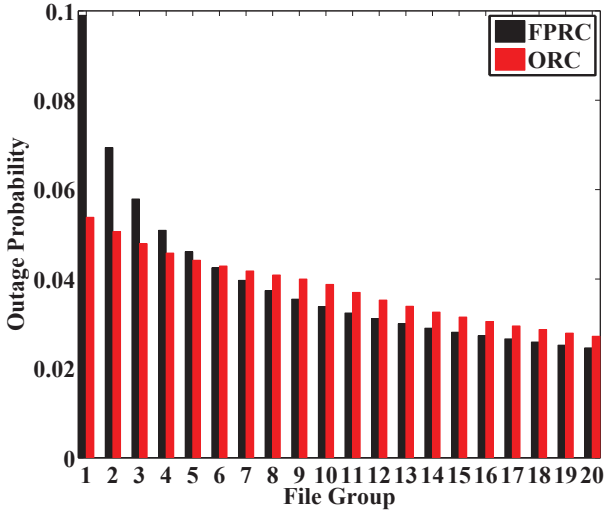}
\caption{Outage probabilities $\Pr(\mathcal Q_n)\cdot P_{\mathcal F_n}$ for individual FGs $\mathcal F_n$ under the file popularity based random caching (FPRC) and optimized random caching (ORC) schemes.}\label{fig:outage_individual}
\end{figure}
\begin{figure}
\centering
\includegraphics[width=\linewidth]{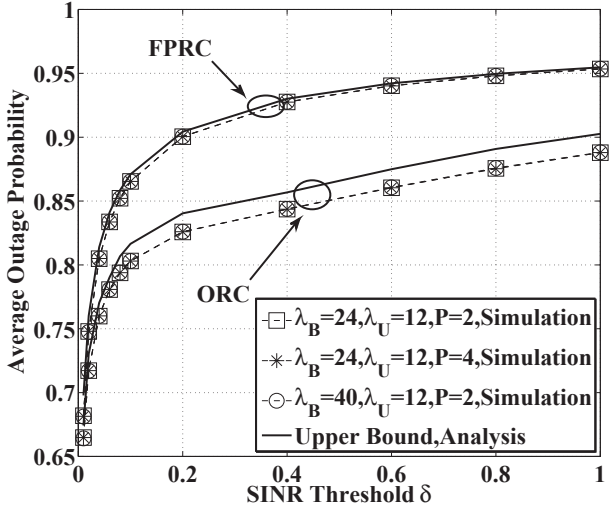}
\caption{Average outage probabilities $\Pr(\mathcal Q)$ vs. $\delta$ under the FPRC and ORC schemes for different SBS and MU intensities $\lambda_B$ and $\lambda_U$, and for transmit powers $P=2$ and $4$.}\label{fig:average_outage_delta}
\end{figure}
\section{Simulation Results}\label{sec:simulations}
In this section, we first focus on the HCNs associated with PPP distributed nodes, where we investigate the average degree distribution of the factor graph and the performance of the random caching scheme. Then we consider an HCN supporting a fixed number of nodes. We investigate the delay optimized by the BP algorithm and compare it to other benchmarks, including both the random caching and the optimal scheme using exhaustive search.

Note that the physical layer parameters in our simulations, such as the path-loss exponent, noise power, transmit power of the SBSs, and the intensity of the SBSs, are chosen to be practical and in line with the values set by 3GPP standards. For instance, the transmit power of an SBS is typically 2 Watt in 3GPP. The unit of power, such as noise power and transmit power, is the classic Watt. The intensities of the SBSs and MUs are expressed in terms of the numbers of the nodes per square kilometer. Unless specified otherwise, we set the path loss to $\alpha=4$, the number of files to $Q=100$, transmit power to $P=2$, and the noise power to $\sigma^2=10^{-10}$. All the simulations are executed with Matlab. Also, we consider the performance averaged over a thousand network cases, where the locations of network nodes are uniformly distributed in each case, and randomly changed from case to case.
\subsection{Average Degree Distributions of Factor Graph}
We compare our Monte-Carlo simulations and analytical results in the HCNs at various transmission powers and node densities. Fig.~\ref{fig:Zeta_U} shows the average degree of the factor nodes with different transmission power $P$, SBSs' intensity $\lambda_B$, and MUs' intensity $\lambda_U$. We can see that for a given $\delta$, the degree $\zeta_U$ remains unaffected by the specific choice of $P$, $\lambda_B$, and $\lambda_U$. Observe that our analytical results are consistent with the simulations. Similarly, Fig.~\ref{fig:Zeta_B} shows the average degree of the variable nodes of different powers and node intensities, demonstrating that the results are independent of the power $P$, but depend on the densities $\lambda_B$ and $\lambda_U$. We can also see that the analytical results match well with the simulation results.
\begin{figure}
\centering
\includegraphics[width=\linewidth]{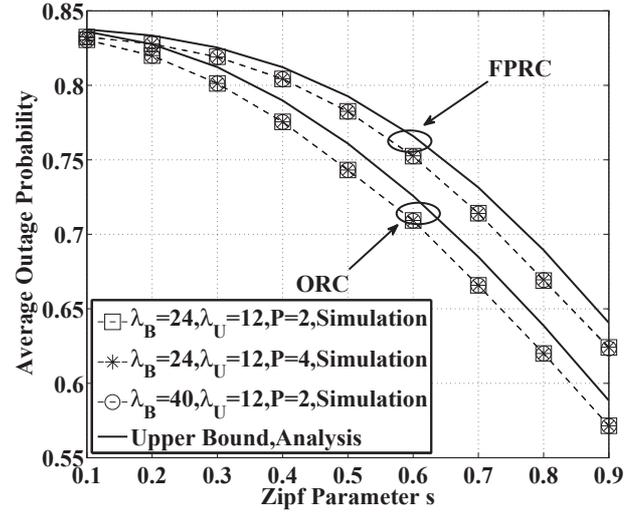}
\caption{Average outage probabilities $\Pr(\mathcal Q)$ vs. the Zipf parameter $s$ under the FPRC and ORC schemes for different SBS and MU intensities $\lambda_B$ and $\lambda_U$, and for transmit powers $P=2$ and $4$.}\label{fig:average_outage_s}
\end{figure}
\begin{figure}
\centering
\includegraphics[width=\linewidth]{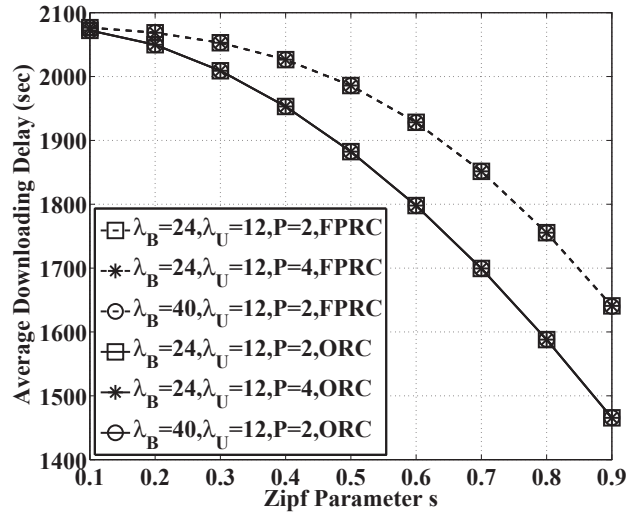}
\caption{Average downloading delay $\bar D$ vs. the Zipf parameter $s$ under the FPRC and ORC schemes for different SBS and MU intensities $\lambda_B$ and $\lambda_U$, and for transmit powers $P=2$ and $4$.}\label{fig:average_delay}
\end{figure}
\subsection{Average Downloading Performance of Random Caching}
Let us now evaluate the average downloading performance of the random caching scheme supporting PPP distributed nodes. The file distribution $\boldsymbol{\mathcal P}=\{p_1,\cdots,p_Q\}$ is modeled by the Zipf distribution~\cite{Cha:2007}, which can be expressed as
\begin{equation}\label{equ:Zipf}
p_f=\frac{1/f^s}{\sum_{q=1}^{Q}1/q^s},\quad\text{for}\;f=1,\cdots,Q,
\end{equation}
where the exponent $0<s\le1$ is a real number, and it characterizes the popularity of files. Explicitly, a larger $s$ corresponds to a higher content reuse, i.e., the most popular files account for the majority of requests. Note that $P_{\mathcal F_n}$ can be obtained based on $p_f$ via Eq.~(\ref{equ:fil_blk_pop}).
\begin{figure}
\centering
\includegraphics[width=\linewidth]{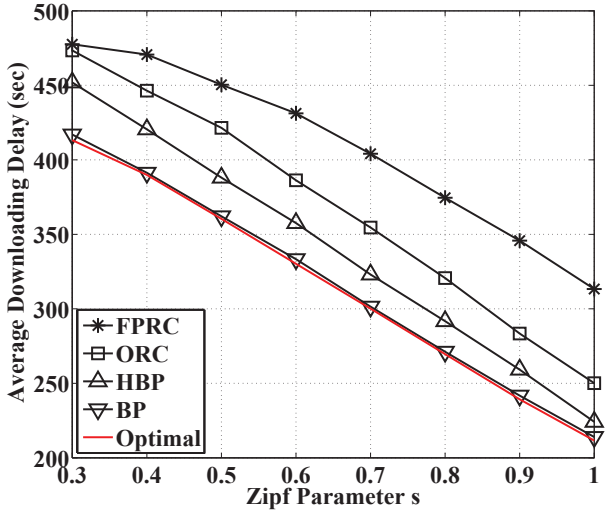}
\caption{Average downloading delay $\bar D$ vs. the Zipf parameter $s$ under various schemes in the first scenario.}\label{fig:Sma_NW}
\end{figure}
\begin{figure}
\centering
\includegraphics[width=\linewidth]{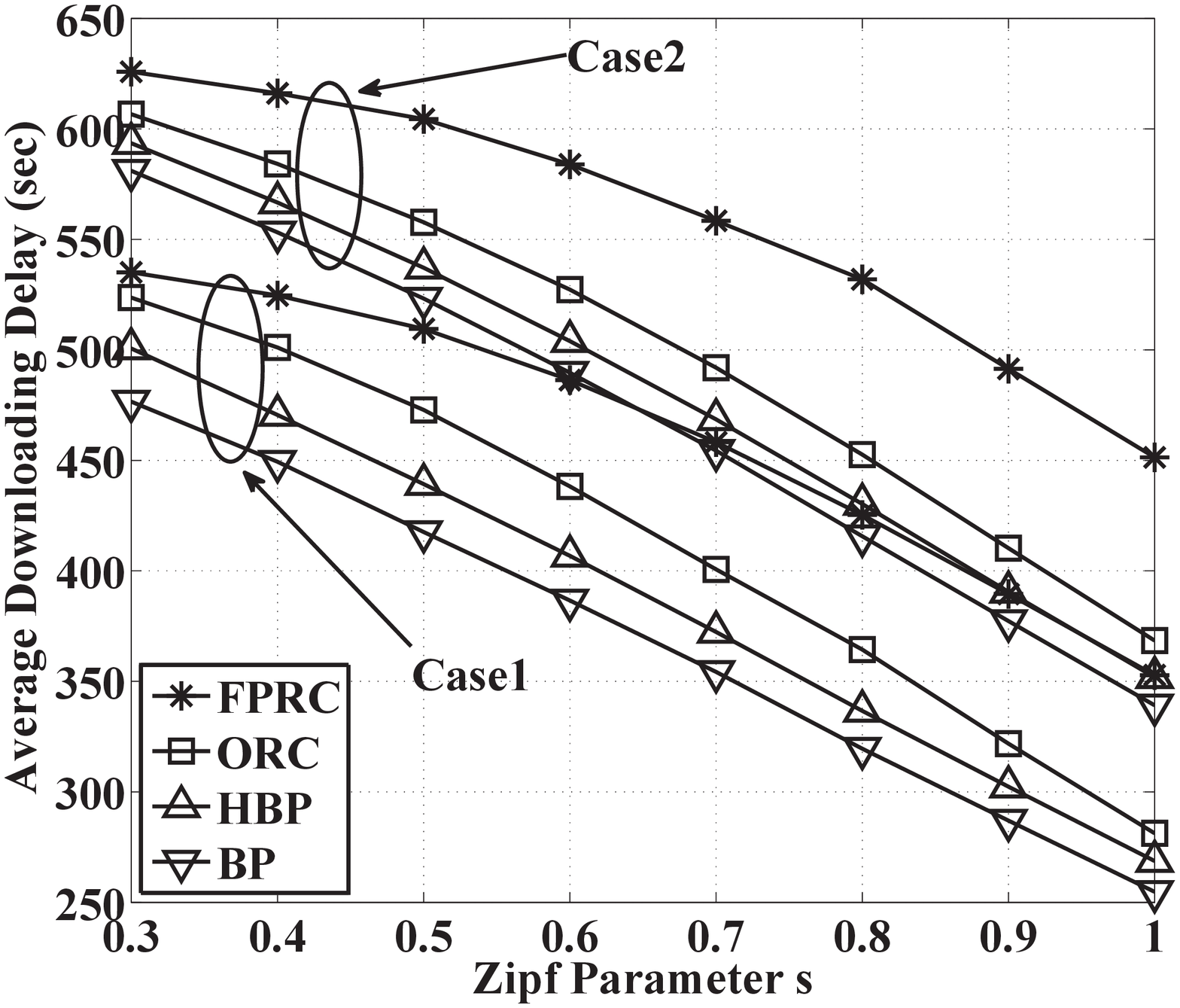}
\caption{Average downloading delay $\bar D$ vs. the Zipf parameter $s$ under various schemes in the second scenario.}\label{fig:Lag_NW}
\end{figure}
\begin{figure}
\centering
\includegraphics[width=\linewidth]{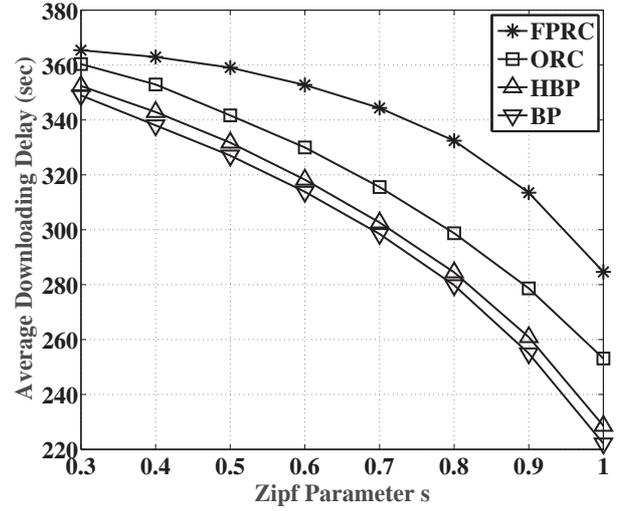}
\caption{Average downloading delay $\bar D$ vs. the Zipf parameter $s$ under various schemes in the large scale network.}\label{fig:LAGS}
\end{figure}

For the simulation results of this subsection, we assume that each SBS caches $G=5$ files, hence there are $N=Q/G=20$ FGs. We commence by considering the OP. In our optimized random caching (ORC), we set $\Omega_{\mathcal F_n}$ as in (\ref{equ:optimal_omega}). For comparison, we also consider another random caching scheme from \cite{Golrezaei:ICC12} as our the benchmark, namely, the file popularity based random caching (FPRC). In the FPRC, $\Omega_{\mathcal F_n}$ is chosen to be consistent with the file popularity, i.e., we have $\Omega_{\mathcal F_n}=P_{\mathcal F_n}$.

Fig.~\ref{fig:outage_individual} shows the OPs $\Pr(\mathcal Q_n)\cdot P_{\mathcal F_n}$ for individual FGs under both the ORC and the FPRC schemes, where we have $\delta=0.03$ and $s=0.5$. The conditional OP $\Pr(\mathcal Q_n)$ (given a file in $\mathcal F_n$ is requested) is calculated from Eq. (\ref{equ:Qn}), while the request probability $P_{\mathcal F_n}$ of $\mathcal F_n$ is calculated from Eq. (\ref{equ:fil_blk_pop}). The FGs are arranged in descending order of popularity, i.e., the first FG has the highest popularity, while the last one has the lowest popularity. We can see from the figure that compared to the FPRC, FGs having a higher popularity have a lower OP, while the ones with lower popularity have higher OPs in the ORC. For example, the OP for the most popular FG is around $0.054$ in the ORC in contrast to $0.099$ in the FPRC, while the probability of the least popular FG is $0.27$ in the ORC in contrast to $0.25$ in the FPRC. This is because the ORC is reminiscent of the classic water-filling, allocating more SBSs for caching the higher popular FGs for ensuring the minimization of the average OP.

Let us now investigate the average OP $\Pr(\mathcal Q)$. Fig.~\ref{fig:average_outage_delta} and Fig. \ref{fig:average_outage_s} show $\Pr(\mathcal Q)$ for different $\delta$ and $s$ values, respectively. In Fig.~\ref{fig:average_outage_delta}, we fix $s=0.5$, while in Fig.~\ref{fig:average_outage_s}, we fix $\delta=0.03$. The dashed lines with different marks are based on the simulations associated with various power and densities, while the solid lines represent the analytical upper bounds of Eq. (\ref{equ:aver_outage}). We can see that the average OP is independent of both the power $P$ and densities $\lambda_B$ and $\lambda_U$. The ORC scheme has a lower average OP than the FPRC. Furthermore, as expected, a higher SINR threshold $\delta$ leads to a higher OP, as shown in Fig.~\ref{fig:average_outage_delta}. At the same time, it is interesting to observe from Fig.~\ref{fig:average_outage_s} that a larger $s$, representing more imbalanced downloading requests on the different files, can dramatically reduce the OP. We can see that the upper bounds evaluated from Eq. (\ref{equ:aver_outage}) match the simulations quite accurately.

Next, we consider the average delay $\bar D$ in Eq. (\ref{equ:downloading_ave}), where we assume an SINR threshold of $\delta=0.03$, a bandwidth of $W=10^7$ Hz, and a file size of $M=10^9$ bits. Since $C_0$ should be always less than the maximum possible downloading rate provided by the SBSs, we assume $C_0=W\log(1+\delta)$. For $\delta=0.03$, $C_0$ becomes $4.26\times 10^5$ bits/sec. Fig.~\ref{fig:average_delay} illustrates the average downloading delay associated with different $s$ values. We can see that the ORC scheme always outperforms the FPRC scheme, and that their performance gap becomes larger upon increasing $s$. Again, the observed performance does not depend on the powers and intensities of the nodes.
\begin{table*}
\caption{The average number of iterations under different $s$.}\label{tab:ite_num}
\centering
\begin{tabular}{|c|c|c|c|c|c|c|c|c|}
\hline
\multicolumn{9}{|c|}{\bfseries Zipf Parameter $s$}\\
\hline
$s$ & $0.3$ & $0.4$ & $0.5$ & $0.6$ & $0.7$ & $0.8$ & $0.9$ & $1.0$\\
\hline
\multicolumn{9}{|c|}{\bfseries Average Number of Iterations for Scenario 1}\\
\hline
$\text{BP}$ & $4.466$ & $4.406$ & $4.002$ & $3.652$ & $3.574$ & $3.412$ & $3.12$ & $2.862$\\
\hline
$\text{HBP}$ & $8.431$ & $8.235$ & $7.634$ & $7.094$ & $6.71$ & $6.494$ & $6.097$ & $5.263$\\
\hline
\multicolumn{9}{|c|}{\bfseries Average Number of Iterations for Scenario 2}\\
\hline
\multicolumn{9}{|c|}{\bfseries Case1}\\
\hline
$\text{BP}$ & $9.429$ & $8.412$ & $7.632$ & $7.326$ & $6.576$ & $5.978$ & $5.804$ & $5.696$\\
\hline
$\text{HBP}$ & $14.973$ & $14.903$ & $14.817$ & $14.783$ & $14.722$ & $14.667$ & $14.623$ & $14.443$\\
\hline
\multicolumn{9}{|c|}{\bfseries Case2}\\
\hline
$\text{BP}$ & $9.548$ & $8.642$ & $7.987$ & $7.483$ & $7.119$ & $6.746$ & $6.057$ & $5.841$\\
\hline
$\text{HBP}$ & $14.994$ & $14.97$ & $14.925$ & $14.821$ & $14.877$ & $14.722$ & $14.648$ & $14.549$\\
\hline
\end{tabular}
\end{table*}
\subsection{Delay Performance of Distributed BP algorithms}
Let us now study the delay performance of distributed BP-based optimizations. We consider HCNs having fixed numbers of SBSs and MUs, where the locations of these nodes are time-variant. We first consider a small network, in which the optimal solution is found with the aid of an exhaustive search. This will allow us to characterize the performance disparity between the proposed BP algorithm and the optimal search-based solution. Then we focus our attention on a larger network to show the robustness of our BP algorithms. In both scenarios, we set the SINR threshold to $\delta=0.1$, the transmission power to $P=2$, the bandwidth to $W=10^7$ Hz, and the file size to $M=10^9$ bits. Similar to the previous subsection, we assume that the rate provided by the MBS as $C_0=W\log(1+\delta)$. For $\delta=0.1$, we have $C_0$ as $1.3\times 10^6$ bits/sec.

In the first scenario, the nodes are arranged in a $0.6\times0.6km^2$ area using $8$ SBSs and $4$ MUs. We assume that each SBS caches $G=25$ files, and there are $N=Q/G=4$ FGs. Fig.~\ref{fig:Sma_NW} shows the average delay performance under various schemes, where `HBP' is the heuristic BP algorithm proposed in Section~\ref{sec:heuristic_BP}, `BP' is the original BP algorithm proposed in Section~\ref{sec:cache_optimizations_BP}, and `Optimal' is the optimal scheme relying on an exhaustive search. We can see from Fig.~\ref{fig:Sma_NW} that the original BP approaches the optimal scheme within a small delay margin. The proposed HBP performs slightly worse than the original BP, with a relatively modest delay degradation of around $5\%$ or $20$ seconds, while it outperforms the ORC scheme by about $10\%$ or $40$ seconds gain. The FPRC performs the worst among all the caching schemes, exhibiting a substantial delay gap between the FPRC scheme and the ORC scheme.

In the second scenario, the nodes are arranged in a $1.5\times1.5km^2$ area with $50$ SBSs and $25$ MUs. We consider two cases, namely Case1 and Case2. In Case1, we assume that each SBS caches $G=20$ files and there are $N=Q/G=5$ FGs, while in Case2, we assume that each SBS caches $G=10$ files and that we have $N=Q/G=10$. Fig.~\ref{fig:Lag_NW} shows the average delay performance under various schemes. It is clear from Fig.~\ref{fig:Lag_NW} that in both cases the BP algorithm performs the best, while the FPRC performs the worst. The HBP exhibits a tiny delay increase of around $3\%$ performance loss compared to the original BP, although it dramatically reduces the communication complexity during the optimization process.

Note also in Fig.~\ref{fig:Lag_NW} that the ORC suffers from a $5\%$ performance loss compared to the HBP, but it is much less complex than the HBP and BP. The optimization in ORC is based on the statistical information available about both of channels and the locations of the nodes, while both the BP and the HBP exploit the relevant instantaneous information at a relatively high communication complexity. In this sense, the ORC constitutes an efficient caching scheme. Furthermore, we can see from Fig.~\ref{fig:Lag_NW} that there is a tradeoff between the storage and delay , i.e., a larger storage at each SBS in Case1 leads to a lower downloading delays compared to Case2.

In the above BP simulations, we set the maximum number of iterations to $T=15$. Table~\ref{tab:ite_num} shows the average number of iterations under different $s$ values for the two scenarios. We can see that the HBP relies on more iterations than the BP. Nevertheless, the overall communication complexity of the HBP is still lower than that of the BP, as we have discussed in Section~\ref{sec:heuristic_BP}. Explicitly, for each iteration of the HBP, ${\mathcal B}_k$ broadcasts $N$ integers and $ {\mathcal U}_j$ transmits $|\boldsymbol{\mathcal H}(j)|$ integers. By contrast, in the original BP, ${\mathcal B}_k$ transmits $N|\boldsymbol{\mathcal H}(k)|$ real numbers and ${\mathcal U}_j$ transmits $N|\boldsymbol{\mathcal H}(j)|$ real numbers.
\subsection{Delay Performance in a Large Scale Network}
Finally, we consider a large-scale network associated with $Q=1000$ files, $50$ SBSs, and $100$ MUs within an area of $5\times 5 km^2$. Furthermore, we consider a lower connection probability to the SBSs by setting $\delta=0.2$. By assuming that each SBS is capable of caching $20$ files, we have overall $50$ file groups. Fig.~\ref{fig:LAGS} shows the average delay performance. We can see from the figure that both BP algorithms perform better than the random caching schemes. Particularly, the HBP has a roughly $1\%$ performance loss compared to the original BP, which imposes however a much reduced communication complexity. This implies that our BP algorithms are robust in large-scale networks associated with a large number of files and network nodes.

Further comparing Figs.~\ref{fig:Sma_NW},~\ref{fig:Lag_NW}, and~\ref{fig:LAGS}, it is interesting to observe that the gap between our BP and HBP algorithms becomes smaller when the network scale becomes larger. More particularly in Fig.~\ref{fig:LAGS}, the performance of these two schemes almost overlaps. This indicate that in large scale networks, we may consider to use the HBP rather than BP to obtain a good performance at a much reduced complexity.
\section{Conclusions}\label{sec:conclusions}
In this paper, we designed distributed caching optimization algorithms with the aid of BP for minimizing the downloading latency in HCNs. Specifically, a distributed BP algorithm was proposed based on the factor graph according to the network structure. We demonstrated that a fixed point of convergence exists for the distributed BP algorithm. Furthermore, we proposed a modified heuristic BP algorithm for further reducing the complexity. To have a better understanding of the average network performance under varying numbers and locations of the network nodes, we involved stochastic geometry theory in our performance analysis. Specifically, we developed the average degree distribution of the factor graph, as well as an upper bound of the OP for random caching schemes. The performance of the random caching was also optimized based on the upper bound derived. Simulations showed that the proposed distributed BP algorithm approaches the optimal performance of the exhaustive search within a small margin, while the modified BP offers a good performance at a very low complexity. Additionally, the average performance obtained by stochastic geometry analysis matches well with our Monte-Carlo simulations, and the optimization based on the upper bound derived provides a better performance than the benchmark of~\cite{Golrezaei:ICC12}.
\appendices
\section{Proof of Lemma~\ref{lem:continous_mapping}}\label{app:continous_mapping}
To simplify the notation in the proof, we assume that $\boldsymbol{\mathcal H}(j)=\boldsymbol{\mathcal K}$, $\forall j\in\boldsymbol{\mathcal J}$ and $\boldsymbol {\mathcal H}(k)=\boldsymbol{\mathcal J}$, $\forall k\in\boldsymbol{\mathcal K}$. Consider a pair of probability vector sets $\boldsymbol{\mathcal M}^{(t-1)}=\left\{p_{k\rightarrow j}^{(t-1)}(\boldsymbol\lambda_k)\right\}$ and $\boldsymbol{\widetilde{\mathcal M}}^{(t-1)}=\left\{\tilde p_{k\rightarrow j}^{(t-1)}(\boldsymbol\lambda_k)\right\}$. Then we have the supremum norm
\begin{equation}\label{equ:two_vectors}
\begin{split}
&\left|\left|\boldsymbol{\Gamma}\left(\boldsymbol{\mathcal M}^{(t-1)}\right)-\boldsymbol{\Gamma}\left(\boldsymbol{\widetilde{\mathcal M}}^{(t-1)}\right)\right|\right|_{\text{sup}}\\&
=\max_{k,j,n}\left|p_{k\rightarrow j}^{(t)}(\boldsymbol\lambda_k^{[n]})-\tilde p_{k\rightarrow j}^{(t)}(\boldsymbol\lambda_k^{[n]})\right|
\\&=\max_{k,j,n}\left|\prod_{i\in{\boldsymbol {\mathcal J}}\backslash\{j\}}\sum_{h\in\boldsymbol{\mathcal K}\backslash\{k\}}\sum_{\boldsymbol\lambda_h=\boldsymbol\lambda_{h}^{[1]}}^{\boldsymbol\lambda_{h}^{[N]}}\left(\exp(\mu F_i(\boldsymbol{\Lambda}_i))
\left(\prod_{q\in\boldsymbol{\mathcal K}\backslash\{k\}}\right.\right.\right.\\&\quad\left.\left.\left.p_{q\rightarrow i}^{(t-1)}(\boldsymbol\lambda_q)
-\prod_{q\in\boldsymbol{\mathcal K}\backslash\{k\}}\tilde p_{q\rightarrow i}^{(t-1)}(\boldsymbol\lambda_q)\right)\right)\right|
\\&\overset{(a)}{\le}\max_{j}\prod_{i\in{\boldsymbol {\mathcal J}}\backslash\{j\}}\sum_{h\in\boldsymbol{\mathcal K}\backslash\{k\}}\sum_{\boldsymbol\lambda_h=\boldsymbol\lambda_{h}^{[1]}}^{\boldsymbol\lambda_{h}^{[N]}}\\&\quad\left|
\prod_{q\in\boldsymbol{\mathcal K}\backslash\{k\}}p_{q\rightarrow i}^{(t-1)}(\boldsymbol\lambda_q)
-\prod_{q\in\boldsymbol{\mathcal K}\backslash\{k\}}\tilde p_{q\rightarrow i}^{(t-1)}(\boldsymbol\lambda_q)\right|
\\&\overset{(b)}{\le}{(K-1){N^{K-1}}}\max_{j}\\&\quad\prod_{i\in{\boldsymbol {\mathcal J}}\backslash\{j\}}\max_{q\in\boldsymbol{\mathcal K}\backslash\{k\},n}\left|
p_{q\rightarrow i}^{(t-1)}(\boldsymbol\lambda_q^{[n]})-\tilde p_{q\rightarrow i}^{(t-1)}(\boldsymbol\lambda_q^{[n]})\right|
\\&\le{(K-1){N^{K-1}}}\max_{j,q\in\boldsymbol{\mathcal K}\backslash\{k\},n}\left|
p_{q\rightarrow i}^{(t-1)}(\boldsymbol\lambda_q^{[n]})-\tilde p_{q\rightarrow i}^{(t-1)}(\boldsymbol\lambda_q^{[n]})\right|^{J-1}
\\&\le{(K-1){N^{K-1}}}\max_{j,k,n}\left|
p_{k\rightarrow i}^{(t-1)}(\boldsymbol\lambda_k^{[n]})-\tilde p_{k\rightarrow i}^{(t-1)}(\boldsymbol\lambda_k^{[n]})\right|
\\&={(K-1){N^{K-1}}}\left|\left|\boldsymbol{\mathcal M}^{(t-1)}-\boldsymbol{\widetilde{\mathcal M}}^{(t-1)}\right|\right|_{\text{sup}}.
\end{split}
\end{equation}

The inequality $(a)$ in (\ref{equ:two_vectors}) is derived by exploiting the following two facts: 1) $0<\exp(\mu F_i(\boldsymbol{\Lambda}))\le 1$, since $F_i(\boldsymbol{\Lambda})$ is non-positive and $\mu$ is positive, and 2) $\sum_s|x_s|\le|\sum_s(x_s)|$ for arbitrary $x_s$. The inequality $(b)$ in (\ref{equ:two_vectors}) can be obtained from: 1) the following lemma, and 2) the fact that $\sum_{h\in\boldsymbol{\mathcal K}\backslash\{k\}}\sum_{\boldsymbol\lambda_h=\boldsymbol\lambda_{h}^{[1]}}^{\boldsymbol\lambda_{h}^{[N]}}$ has to carry out the additions of $N^{K-1}$ items.
\begin{lemma}\label{lem:inequality_b}
Given $0\le a_1,\cdots,a_K\le1$ and $0\le\tilde a_1,\cdots,\tilde a_K\le1$, we have
\begin{equation}\label{equ:ineq_lemma}
\max_{k\in{\boldsymbol{\mathcal K}}}\left|\prod_{q\in{\boldsymbol {\mathcal K}\backslash\{k\}}} a_q-\prod_{q\in{\boldsymbol {\mathcal K}\backslash\{k\}}} \tilde a_q\right|\le (K-1)\max_{q\in\boldsymbol {\mathcal K}\backslash\{k\}}|a_q-\tilde a_q|.
\end{equation}

\emph{Proof:} Please refer to Appendix~\ref{app:inequality_b}.
\end{lemma}

From (\ref{equ:two_vectors}), we can infer that $\boldsymbol{\Gamma}$ is a continuous mapping, since the coefficient ${(K-1){N^{K-1}}}$ is a constant, and this completes the proof.
\hfill$\blacksquare$
\section{Proof of Theorem~\ref{theo:fixed_point}}\label{app:fixed_point}
Let $\boldsymbol{\mathcal S}$ be the collection of the message set $\boldsymbol{\mathcal M}^{(t)}$. The mapping function $\boldsymbol {\Theta}$ maps $\boldsymbol{\mathcal S}$ to $\boldsymbol{\mathcal S}$ with the aid of the function $\boldsymbol {\Gamma}$. According to Lemma \ref{lem:continous_mapping}, $\boldsymbol {\Theta}$ is continuous since $\boldsymbol {\Gamma}$ is continuous. Furthermore, it is clear that the set $\boldsymbol{\mathcal S}$ is convex, closed and bounded. Based on Schauder's fixed point theorem, $\boldsymbol {\Theta}$ has a fixed point. This completes the proof.\hfill$\blacksquare$
\section{Proof of Theorem~\ref{theo:average}}\label{app:average}
\subsection{The Average Degree of Factor Nodes}
Without a loss of generality, we carry out the analysis for a typical MU located at the origin and assume that the potential serving SBSs are located at the point $x_{B}$. The fading (power) is denoted by $h_{x_{B}}$, which is assumed to be exponentially distributed, i.e., we have $h_{x_B}\sim \exp(1)$. The path-loss function is given by ${\left\Vert x_{B}\right\Vert}^{-\alpha}$, where $\Vert\cdot\Vert$ denotes the Euclidian distance.

The average degree of a factor node in the factor graph is equivalent to the number of SBSs that can provide a high enough SINR ($\ge \delta$) for the typical MU, which can be formulated as
\begin{equation}\label{equ:Nl}
N_{B}=\int_{\mathbb{R}^{2}}\lambda_{B}\Pr\left(\rho(x_{B})\ge\delta\right)\text{d}x_{B},
\end{equation}
where $\rho(x_{B})$ represents the SINR at the typical MU received from the SBSs located at $x_{B}$.

We first focus on the probability $\Pr\left(\rho(x_{B})\ge\delta\right)$ in (\ref{equ:Nl}) as follows.
\begin{align}\label{Eq:SINR_Threshold}
\Pr\left(\rho(x_{B})\ge\delta\right)&=\Pr\left(\frac{Ph_{x_{B}}\left\Vert x_{B}\right\Vert ^{-\alpha}}{\underset{x_{k} \in\Phi_{B}}{\sum}Ph_{x_{k}}\left\Vert x_{k}\right\Vert ^{-\alpha}+\sigma^{2}}\ge\delta \right)  \nonumber \\
&=\Pr\left(h_{x_{B}}\ge\frac{\delta\left(I+\sigma^{2}\right)}{P\left\Vert x_{B}\right\Vert ^{-\alpha}} \right) \nonumber \\
&=\mathbb{E}_{I}\left(\exp\left(-sI\right)\right)\exp\left(-s\sigma^{2}\right),
\end{align}
where $x_{k}$ denotes the location of an interfering SBS, $I\overset{\triangle}{=}\underset{x_{k} \in\Phi_{B}}{\sum}Ph_{x_{k}}\left\Vert x_{k}\right\Vert ^{-\alpha}$ represents the aggregate interference, and $s=\frac{\delta\left\Vert x_{B}\right\Vert ^{\alpha}}{P}$. The last step is due to the exponential distribution of $h_{x_{B}}$. Then, we derive $\mathbb{E}_{I}\left(\exp\left(-sI\right)\right)$ in (\ref{Eq:SINR_Threshold}) as
\begin{equation}\label{Eq:Interference}
\begin{split}
&\mathbb{E}_{I}\left(\exp\left(-sI\right)\right)\overset{\left(a\right)}{=} \\&\mathbb{E}_{\Phi_{B}}\left(\underset{x_{k} \in\Phi_{B}}{\prod} \int_{0}^{\infty}\exp\left(-sPh_{x_{k}}\left\Vert x_{k}\right\Vert ^{-\alpha}\right)\exp(-h_{x_{k}})\text{d}h_{x_{k}}\right)   \\&
\overset{\left(b\right)}{=}\exp\left(-\lambda_{B}\int_{\mathbb{R}^{2}}\left(1-\frac{1}{1+sP\left\Vert x_{k}\right\Vert ^{-\alpha}}\right)\text{d}x_{k}\right)  \\&
=\exp\left( -2\pi\lambda_{B}\frac{1}{\alpha}\left(sP\right)^{\frac{2}{\alpha}}B\left(\frac{2}{\alpha},1-\frac{2}{\alpha}\right)\right),
\end{split}
\end{equation}
where $(a)$ is based on the independence of channel fading, and $(b)$ follows from $\mathbb{E}\left(\underset{x}{\prod}u\left(x\right)\right)=\exp\left(-\lambda\int_{\mathbb{R}^{2}}\left(1-u\left(x\right)\right)\text{d}x\right)$, where $x\in \Phi$ and $\Phi$ is an PPP in $\mathbb{R}^{2}$ with the intensity $\lambda$ \cite{Daley:book}.

Based on the derivation above, the average degree of the typical MU can be calculated as
\begin{equation}\label{equ:NB}
\begin{split}
&N_B=\lambda_{B}\int_{\mathbb{R}^{2}}\\&\exp\left(-2\pi\frac{\lambda_{B}}{\alpha}{\delta}^{\frac{2}{\alpha}}B\left(\frac{2}{\alpha},1-\frac{2}{\alpha}\right)\left\Vert x_{B}\right\Vert ^{2}-\frac{\delta\sigma^{2}}{P}\left\Vert x_{B}\right\Vert ^{\alpha}\right)\text{d}x_{B} \\& =2\pi\lambda_{B}\int^{\infty}_{0}\\&\exp\left(-2\pi\frac{\lambda_{B}}{\alpha}\delta^{\frac{2}{\alpha}}B\left(\frac{2}{\alpha},1-\frac{2}{\alpha}\right)r^{2}-\frac{\delta\sigma^{2}}{P}r^{\alpha}\right)r\text{d}r.
\end{split}
\end{equation}
\subsection{The Average Degree of Variable Nodes}
In this subsection, we consider a typical SBS which is located at the origin, and assume that an MU is located at the point $x_U$. The average degree of a variable node in the factor graph is equivalent to the number of MUs that can receive at a high enough SINR ($\ge \delta$) from the typical SBS, which can be formulated as
\begin{equation}
N_{U}=\int_{\mathbb{R}^{2}}\lambda_{U}\Pr\left(\rho(x_U)\ge\delta\right)\text{d}x_U,
\end{equation}
where $\rho(x_U)$ represents the received SINR at the MU located at $x_U$ from the typical SBS, i.e.,
\begin{multline}
\Pr\left(\rho(x_U)\ge\delta\right)\\=\Pr\left(\frac{Ph_{x_U}\left\Vert x_U\right\Vert ^{-\alpha}}{\underset{x_{k} \in\Phi_{B}}{\sum}Ph_{x_k}\left\Vert x_{k}-x_U\right\Vert ^{-\alpha}+\sigma^{2}}\ge\delta \right),
\end{multline}
where $x_{k}$ denotes the location of an interfering SBS.

Since the PPP is a stationary process, the distribution of $\left\Vert x_{k}-x_U\right\Vert$ is independent of the value of $x_U$, i.e., we have $p(\left\Vert x_{k}-x_U\right\Vert)=p(\left\Vert x_{k}\right\Vert)$, where $p(\cdot)$ represents the probability density function. Then, we have similar results to Eq. (\ref{Eq:SINR_Threshold}). That is, we have
\begin{equation}\label{equ:PRXU}
\Pr\left(\rho(x_U)>\delta\right)=\mathbb{E}_{I}\left(\exp\left(-sI\right)\right)\exp\left(-s\sigma^{2}\right),
\end{equation}
where $s=\frac{\delta\left\Vert x_U\right\Vert ^{\alpha}}{P}$. Then we arrive at
\begin{multline}\label{equ:NU}
N_U=2\pi\lambda_{U}\\\int^{\infty}_{0}\exp\left(-2\pi\frac{\lambda_{B}}{\alpha}\delta ^{\frac{2}{\alpha}}B\left(\frac{2}{\alpha},1-\frac{2}{\alpha}\right)r^{2}-\frac{\delta\sigma^{2}}{P}r^{\alpha}\right)r\text{d}r.
\end{multline}
By combining Eqs. (\ref{equ:NU}) and (\ref{equ:NB}), we complete the proof.\hfill$\blacksquare$
\section{Proof of Corollary~\ref{cor:alpha}}\label{app:MU_nolambda}
When ignoring the noise, we have
\begin{equation}
\begin{split}
&Z(\lambda_B, P, \alpha, \delta)=\int^{\infty}_{0}\exp\left(-\frac{2\pi\lambda_{B}}{\alpha}\delta^{\frac{2}{\alpha}}B\left(\frac{2}{\alpha},1-\frac{2}{\alpha}\right)r^{2}\right)r\text{d}r \\
&=\frac{1}{2}\int_{0}^{\infty}\exp\left(-\lambda_{B}\frac{2\pi}{\alpha}\delta^{\frac{2}{\alpha}}B\left(\frac{2}{\alpha},1-\frac{2}{\alpha}\right)t\right)\text{d}t \\&=\frac{1}{2\lambda_{B}\frac{2\pi}{\alpha}\delta^{\frac{2}{\alpha}}B\left(\frac{2}{\alpha},1-\frac{2}{\alpha}\right)}=\frac{\alpha}{4\pi\lambda_B B\left(\frac{2}{\alpha},1-\frac{2}{\alpha}\right) \delta^{\frac{2}{\alpha}}}.
\end{split}
\end{equation}
By substituting the above expression into (\ref{Eq:average_BS}) and (\ref{Eq:average_MU}), we obtain (\ref{equ:approx_du}) and (\ref{equ:approx_dl}) respectively. This completes the proof.\hfill$\blacksquare$
\section{Proof of Theorem~\ref{theo:outage_prob}}\label{app:outage_prob}
We conduct the analysis for a typical MU that is located at the origin. We assume that when downloading a file in $\mathcal F_n$, the MU will always associate with its nearest SBS, which caches $\mathcal F_n$. Note that the OP derived under this assumption is an upper bound for the exact OP. This is because the MU will associate with the second-nearest SBS if it can provide a higher received SINR than that provided by the nearest SBS. Therefore, in some cases, the nearest SBS cannot provide a higher enough SINR ($\ge\delta$), while the second-nearest SBS can. According to our assumption, we will neglect these cases, which leads to a higher OP.

Let us denote by $z$ the distance between the typical MU and the nearest SBS that caches $\mathcal F_n$. The location of the nearest SBS caching $\mathcal F_n$ is denoted by $x_Z$. The fading (power) for an SBS located at $x_B$, $\forall x_B\in \Phi_{B}$, is denoted by $h_{x_B}$, which is assumed to be exponentially distributed, i.e., $h_{x_B}\sim \exp(1)$. The path-loss function for a given point $x_B$ is ${\left\Vert x_B\right\Vert}^{-\alpha}$.

When random caching is adopted, the distribution of the SBSs that cache $\mathcal F_n$ can be modeled as an PPP with the intensity of $\Omega_{\mathcal F_n}\lambda_B$. The event that the typical MU can download a file in $\mathcal F_n$ from an SBS means that the received SINR from the nearest SBS which caches $\mathcal F_n$ is no less than the threshold $\delta$. Let us denote by $\rho(x_Z)$ the received SINR at the typical MU from the nearest SBS. Then the average probability that the MU can download the file from an SBS is
\begin{equation}\label{eq:probability_femto}
\begin{split}
&\Pr(\rho(x_Z)\ge\delta)\\&=\int_{0}^{\infty}\Pr\left(\left.\frac{h_{x_Z}z^{-\alpha}}{\underset{x_k\in\Phi_{B}\backslash\{x_Z\}}{\sum}h_{x_k}\left\Vert x_k\right\Vert ^{-\alpha}}\ge\delta\right| z\right)f_{Z}\left(z\right)\text{d}z \\
&=\int_{0}^{\infty}\Pr\left(\left.h_{x_Z}\ge\frac{\delta\left(\underset{x_k\in\Phi_{B}\backslash\{x_Z\}}{\sum}h_{x_k}\left\Vert x_k\right\Vert ^{-\alpha}\right)}{z^{-\alpha}}\right|z\right)\cdot\\&\qquad 2\pi \Omega_{\mathcal F_n}\lambda_B z \exp(-\pi \Omega_{\mathcal F_n}\lambda_B z^2)\:\text{d}z \\
&=\int_{0}^{\infty}\mathbb{E}_{I}\left(\exp\left(-z^{\alpha}\delta I\right)\right)2\pi \Omega_{\mathcal F_n}\lambda_B z\exp(-\pi \Omega_{\mathcal F_n}\lambda_B z^2)\:\text{d}z,
\end{split}
\end{equation}
where we have $I\triangleq\underset{x_k\in\Phi_{B}\backslash\{x_Z\}}{\sum}h_{x_k}\left\Vert x_k\right\Vert ^{-\alpha}$, and the PDF of $z$, i.e., $f_{Z}\left(z\right)$, is derived by the null probability of a Poisson process with the intensity of $\Omega_{\mathcal F_n}\lambda_B$. Note that the interference $I$ consists of $I_1$ and $I_2$, where $I_1$ is emanating from the SBSs caching the FGs $\mathcal F_q$, $\forall q\in{\boldsymbol {\mathcal N}},q\neq n$, while $I_2$ is from the SBSs caching $\mathcal F_n$ excluding $x_Z$. The SBSs contributing to $I_1$, denoted by $\Phi_{\bar n}$, have the intensity $(1-\Omega_{\mathcal F_n})\lambda_B$, while those contributing to $I_2$, denoted by $\Phi_{n}$, have the intensity $\Omega_{\mathcal F_n}\lambda_B$. Correspondingly, the calculation of $\mathbb{E}_{I}\left(\exp\left(-z^{\alpha}\delta I\right)\right)$ will be split into the product of two expectations over $I_1$ and $I_2$. The expectation over $I_1$ directly follows (\ref{Eq:Interference}), i.e., we have
\begin{equation}\label{equ:expect_I_1}
\mathbb{E}_{I_1}\left(\exp\left(-z^{\alpha}\delta I_1\right)\right)=\exp\left(-\pi(1-\Omega_{\mathcal F_n})\lambda_B C(\delta,\alpha)z^2\right),
\end{equation}
where $C(\delta,\alpha)$ has been defined as $\frac{2}{\alpha}\delta^{\frac{2}{\alpha}}B\left(\frac{2}{\alpha},1-\frac{2}{\alpha}\right)$. The expectation over $I_2$ has to take into account $z$ as the distance from the nearest interfering SBS, i.e., we obtain
\begin{equation}\label{equ:expect_I_2}
\begin{split}
&\mathbb{E}_{I_2}\left(\exp(-z^{\alpha}\delta I_2)\right)\\&=\exp\left(-\Omega_{\mathcal F_n}\lambda_B 2\pi\int_{z}^{\infty}\left(1-\frac{1}{1+ z^{\alpha}\delta r^{-\alpha}}\right)r\text{d}r\right)\\
&\overset{(a)}{=}\exp\left(-\Omega_{\mathcal F_n}\lambda_B \pi\delta^{\frac{2}{\alpha}}z^{2}\frac{2}{\alpha}\int_{\delta^{-1}}^{\infty}\frac{x^{\frac{2}{\alpha}-1}}{1+x}\;\text{d}x\right) \\
&\overset{(b)}{=}\exp\left(-\Omega_{\mathcal F_n}\lambda_B\pi\delta z^{2}\frac{2}{\alpha-2}\; {{}_2}F_{1}\left(1,1-\frac{2}{\alpha};2-\frac{2}{\alpha};-\delta\right)\right),
\end{split}
\end{equation}
where $\left(a\right)$ defines $x\triangleq\delta^{-1}z^{-\alpha}r^{\alpha}$, and ${{}_2}F_{1}(\cdot)$ in $(b)$ is the hypergeometric function. Since we have defined $\emph{A}(\delta,\alpha)=\frac{2\delta}{\alpha-2}\; {{}_2}F_{1}\left(1,1-\frac{2}{\alpha};2-\frac{2}{\alpha};-\delta\right)$, by substituting (\ref{equ:expect_I_1}) and (\ref{equ:expect_I_2}) into (\ref{eq:probability_femto}), we have
\begin{equation}\label{equ:Pr_xZ}
\begin{split}
&\Pr(\rho(x_Z)\ge\delta)=\int_{0}^{\infty}\exp\left(-\pi(1-\Omega_{\mathcal F_n})\lambda_B C(\delta,\alpha)z^2\right)\\&\exp\left(-\pi\Omega_{\mathcal F_n}\lambda_B z^{2}\emph{A}(\delta,\alpha)\right)2\pi \Omega_{\mathcal F_n}\lambda_B z\exp\left(-\pi \Omega_{\mathcal F_n}\lambda_B  z^{2}\right)\text{d}z \\&=\frac{\Omega_{\mathcal F_n}}{C(\delta, \alpha)(1-\Omega_{\mathcal F_n})+\emph{A}(\delta,\alpha)\Omega_{\mathcal F_n}+\Omega_{\mathcal F_n}}.
\end{split}
\end{equation}
It is clear that $\Pr(\mathcal Q_n)=1-\Pr(\rho(z)\ge\delta)$. This completes the proof.
\hfill$\blacksquare$
\section{Proof of Lemma~\ref{lem:inequality_b}}\label{app:inequality_b}
Without loss of generality, we assume $k=1$. Then (\ref{equ:ineq_lemma}) becomes
\begin{equation}\label{equ:convert_inequ}
\left|\prod_{q=2}^{K} a_q-\prod_{q=2}^{K} \tilde a_q\right|\le(K-1) \max_{q\in\{2,\cdots,K\}}|a_q-\tilde a_q|.
\end{equation}
Again, without loss of generality, we assume
\begin{equation}\label{equ:largest_a}
|a_2-\tilde a_2|\ge\cdots\ge|a_K-\tilde a_K|.
\end{equation}

First, we prove that $|a_{K-1}a_K-\tilde a_{K-1}\tilde a_K|\le 2|a_{K-1}-\tilde a_{K-1}|$, under the condition of $|a_{K-1}-\tilde a_{K-1}|\ge|a_K-\tilde a_K|$. To prove this, we discuss the following possible cases.
\subsubsection{When $a_{K-1}\ge \tilde a_{K-1}$ and $a_K\ge\tilde a_K$}
We have $a_K\le a_{K-1}-\tilde a_{K-1}+\tilde a_K$. Then
\begin{equation}\label{equ:case1}
\begin{split}
&|a_{K-1} a_K-\tilde a_{K-1}\tilde a_K|\\&\qquad\le|a_{K-1}(a_{K-1}-\tilde a_{K-1}+\tilde a_K)-\tilde a_{K-1}\tilde a_K|\\&\qquad=|(a_{K-1}+\tilde a_K)(a_{K-1}-\tilde a_{K-1})|\\&\qquad\le 2|a_{K-1}-\tilde a_{K-1}|.
\end{split}
\end{equation}
\subsubsection{When $a_{K-1}\ge \tilde a_{K-1}$, $a_K\le\tilde a_K$, and $a_{K-1}a_K\ge\tilde a_{K-1}\tilde a_K$}
We have
\begin{equation}\label{equ:case2}
\begin{split}
&|a_{K-1}a_{K}-\tilde a_{K-1}\tilde a_{K}|\le|a_{K-1}\tilde a_{K}-\tilde a_{K-1}\tilde a_{K}|\\&\qquad\qquad=|a_{K-1}-\tilde a_{K-1}|\tilde a_{K}\le |a_{K-1}-\tilde a_{K-1}|.
\end{split}
\end{equation}
\subsubsection{When $a_{K-1}\ge \tilde a_{K-1}$, $a_K\le\tilde a_K$, and $a_{K-1}a_K\le\tilde a_{K-1}\tilde a_K$}
We have
\begin{equation}\label{equ:case3}
\begin{split}
&|\tilde a_{K-1}\tilde a_K-a_{K-1}a_K|\le|a_{K-1}\tilde a_K-a_{K-1}a_K|\\&\qquad\qquad=|a_K-\tilde a_K|a_{K-1}\le |a_{K-1}-\tilde a_{K-1}|.
\end{split}
\end{equation}
\subsubsection{When $a_{K-1}\le \tilde a_{K-1}$, $a_K\ge\tilde a_K$, and $a_{K-1}a_K\ge\tilde a_{K-1}\tilde a_K$}
We have
\begin{equation}\label{equ:case4}
\begin{split}
&|a_{K-1}a_K-\tilde a_{K-1}\tilde a_K|\le|\tilde a_{K-1} a_K-\tilde a_{K-1}\tilde a_K|\\&\qquad\qquad=|a_K-\tilde a_K|\tilde a_{K-1}\le |a_{K-1}-\tilde a_{K-1}|.
\end{split}
\end{equation}
\subsubsection{When $a_{K-1}\le \tilde a_{K-1}$, $a_K\ge\tilde a_K$, and $a_{K-1}a_K\le\tilde a_{K-1}\tilde a_K$}
We have
\begin{equation}\label{equ:case5}
\begin{split}
&|\tilde a_{K-1}\tilde a_K-a_{K-1}a_K|\le|\tilde a_{K-1} a_K-a_{K-1}a_K|\\&\qquad\qquad=|a_{K-1}-\tilde a_{K-1}|a_K\le |a_{K-1}-\tilde a_{K-1}|.
\end{split}
\end{equation}
\subsubsection{When $a_{K-1}\le \tilde a_{K-1}$, $a_K\le\tilde a_K$} We have $a_K\ge \tilde a_K+a_{K-1}-\tilde a_{K-1}$. Then
\begin{equation}\label{equ:case6}
\begin{split}
&|\tilde a_{K-1}\tilde a_K-a_{K-1}a_K|\\&\qquad\le|\tilde a_{K-1} \tilde a_K-a_{K-1}(\tilde a_K+a_{K-1}-\tilde a_{K-1})|\\&\qquad=|(a_{K-1}+\tilde a_K)(\tilde a_{K-1}-a_{K-1})|\\&\qquad\le 2|a_{K-1}-\tilde a_{K-1}|.
\end{split}
\end{equation}
From the above discussions, we can see that $|a_{K-1}a_K-\tilde a_{K-1}\tilde a_K|\le 2|a_{K-1}-\tilde a_{K-1}|$.

Second, as there is $|a_{K-1}a_K-\tilde a_{K-1}\tilde a_K|\le 2|a_{K-1}-\tilde a_{K-1}|$, we have $|a_{K-1}a_K-\tilde a_{K-1}\tilde a_K|\le 2|a_{K-2}-\tilde a_{K-2}|$. With this condition, we can prove that $|a_{K-2}a_{K-1}a_K-\tilde a_{K-2}\tilde a_{K-1}\tilde a_K|\le 3|a_{K-2}-\tilde a_{K-2}|$ by following the similar steps above. By doing this iteratively, we have
\begin{equation}\label{equ:convert_inequ_final}
\left|\prod_{q=2}^{K} a_q-\prod_{q=2}^{K} \tilde a_q\right|\le(K-1) |a_2-\tilde a_2|.
\end{equation}
This completes the proof.\hfill$\blacksquare$

\bibliographystyle{IEEEtran}
\bibliography{IEEEabrv,Caching}

\end{document}